%% file: main.tex
\documentclass[aip,amsmath,amssymb,reprint]{revtex4-1}

\usepackage{graphicx}% Include figure files
\usepackage{bm}% bold math

\usepackage{dcolumn}
\usepackage{inputenc}
\usepackage[T1]{fontenc}
\usepackage{mathptmx}
\usepackage{etoolbox}
\usepackage{xcolor}
\usepackage{tabularx}
\usepackage{enumitem}
\usepackage{float}
\usepackage[ruled,vlined]{algorithm2e}

\usepackage{subcaption} %for subfigures
\usepackage{hyperref}
\usepackage{calc}
\usepackage{booktabs}
\usepackage{siunitx}
\usepackage{array}

\usepackage{amsfonts, amsthm, amssymb, amsmath}	
\usepackage{color}
\usepackage{graphicx}
\usepackage{mathrsfs}
\usepackage[english]{babel}
\usepackage{ upgreek }
\usepackage{ stmaryrd }
\usepackage{pgf,tikz}

\theoremstyle{plain}% Theorem-like structures
\newtheorem{Th}{Theorem}[section]
\newtheorem{Proposition}[Th]{Proposition}

\theoremstyle{remark}
\newtheorem*{Note}{Remark}

\usepackage{tikz}
\usetikzlibrary{arrows.meta}
\usetikzlibrary{calc}

\begin{document}

\title[Low-rank matrix and tensor approximations for compression of machine-learning interatomic potentials]{Low-rank matrix and tensor approximations for compression of machine-learning interatomic potentials}

\author{Igor Vorotnikov}
\affiliation{HSE University, Faculty of Computer Science, Pokrovsky boulevard 11, Moscow, 109028, Russian Federation}

\author{Fedor Romashov}
\affiliation{HSE University, Faculty of Computer Science, Pokrovsky boulevard 11, Moscow, 109028, Russian Federation}

\author{Nikita Rybin}
\affiliation{Skolkovo Institute of Science and Technology, Skolkovo Innovation Center, Bolshoy boulevard 30, Moscow, 143026, Russian Federation}
\affiliation{Digital Materials LLC, Odintsovo, Kutuzovskaya str. 4A Moscow region, 143001, Russian Federation}

\author{Maxim Rakhuba}
\affiliation{HSE University, Faculty of Computer Science, Pokrovsky boulevard 11, Moscow, 109028, Russian Federation}

\author{Ivan S. Novikov}
\email{ivan.novikov0590@gmail.com}
\affiliation{HSE University, Faculty of Computer Science, Pokrovsky boulevard 11, Moscow, 109028, Russian Federation}

\date{\today}

\begin{abstract}

Machine-learning interatomic potentials (MLIPs) have become a mainstay in computationally-guided materials science, surpassing traditional force fields due to their flexible functional form and superior accuracy in reproducing physical properties of materials. This flexibility is achieved through mathematically-rigorous basis sets that describe interatomic interactions within a local atomic environment. The number of parameters in these basis sets influences both the size of the training dataset required and the computational speed of the MLIP. Consequently, compressing MLIPs by reducing the number of parameters is a promising route to more efficient simulations. In this work, we use low-rank matrix and tensor factorizations under fixed-rank constraints to achieve this compression. In addition, we demonstrate that an algorithm with automatic rank augmentation helps to find a deeper local minimum of the fitted potential. The methodology is mainly verified using the Moment Tensor Potential (MTP) model and benchmarked on multi-component systems: a Mo-Nb-Ta-W medium-entropy alloy, molten LiF-NaF-KF, and a glycine molecular crystal. The proposed approach achieves up to 50\% compression without any loss of MTP accuracy. We also demonstrate that the developed methodology is universal and can be applied to compress other MLIPs on the example of Atomic Cluster Expansion (ACE). 

\end{abstract}

\maketitle

\section{Introduction}

Nowadays, machine-learning interatomic potentials (MLIPs) are ubiquitously used in computational materials science \cite{friederich2021machine}. The flexible functional form of MLIPs enables to approximate the potential energy surface on which atoms move with arbitrary accuracy \cite{eyert2023_recent_MLIPs}. The potentials are usually parametrized on the training data obtained in highly accurate, but computationally demanding, first principles calculations performed in the scope of density functional theory (DFT). Trained MLIPs reproduce physical behavior of materials and overcome the length- and time-scale limitations of computationally expensive DFT calculations, which made MLIPs an attractive tool in atomistic modeling.

Many MLIPs have been developed since 2007. The first of them are Neural Network Potential (NNP) \cite{behler2007-NNP} based on artificial neural networks, Gaussian Approximation Potential (GAP) \cite{bartok2010-GAP} based on Gaussian processes, and the polynomial-like Spectral Neighbor Analysis Potential (SNAP) \cite{thompson2015-automated} based on spherical harmonics and Moment Tensor Potential (MTP) \cite{shapeev2016-mtp} based on the tensors of inertia of atomistic environments. Most of recently developed MLIPs are based on neural networks \cite{wang2018-deepmd,pun2019-pinn,batzner2022-equivarnn,batatia2022_mace} and some of them are polynomial-like \cite{drautz2019-ACE,hodapp2024-ETN}. In Ref. \cite{deringer2019_MLIP,zhang2025_roadmap}, success of applying MLIPs to materials science problems and challenges in their future development are discussed. 

One of the drawbacks of modern MLIPs is the large number of parameters in their functional form. For example, in Ref. \cite{batatia2022_mace}, three message-passing MLIPs  --- BOTNet, NequIP\cite{batzner2022-equivarnn}, and MACE\cite{batatia2022_mace} ---  were fitted on the rMD17, 3BPA, and AcAc benchmark datasets, including about several hundred thousand atomic structures. Low fitting errors (up to several meV/atom) were obtained on these training sets. However, these models have about 3 million parameters each. There is a very well-known fact that for reaching high accuracy it is necessary to have many parameters in machine learning models, but they require a huge amount of data to avoid overfitting. Therefore, methods for compression, or, reducing parameters in MLIPs enables for preserving accuracy of original (non-compressed) MLIPs become a promising tool and a key to simulate atomistic systems more efficiently and to reduce a number of expensive DFT calculations.

The low-rank factorization approach provides an efficient method for compressing matrices and tensors. This technique has found significant utility in machine learning, especially in deep learning, where it is applied to decompose large weight matrices and tensors to reduce model complexity~\cite{lebedev2014speeding,novikov2015tensorizing,hsu2022language,wang2024svd,mo2025parameter}. 
For example, Ref.~\cite{mo2025parameter} recently used low-rank matrix decomposition to pre-train a large language model, achieving a 54\% reduction in memory usage while also improving its perplexity score. 

Inspired by the success of applying low-rank approximations to neural networks, in this study, we introduce and apply methods for compression of MLIPs. In this framework, low-rank matrix and tensor decompositions (or factorizations)~\cite{kolda2009tensor,grasedyck2013literature,khoromskij2018tensor} are used to reduce a number of MLIP parameters without losing accuracy of its fitting. We divide the methods for compression of MLIPs into two types. The first type is optimization under fixed-rank constraints. We apply fixed-rank matrix and tensor factorizations to MLIP parameters and reduce their number. We explore both standard optimization techniques for the parameters of fixed-rank matrix/tensor factorization and the Riemannian optimization~\cite{absil2009optimization,uschmajew2020geometric,qi2010riemannian} approach, which additionally removes excessive parameters arising in rank factorizations. The second type of methods proposed in this study is the rank augmentation method. Here, we start with a small rank during compressed MLIP fitting and increase it to a specific value. Once this value is reached, we continue the fitting with this rank. We use this method to fit MLIP compressed with matrix factorization.

In this work, our aim is to verify that compressed MLIPs are as accurate as the base (non-compressed) ones. Hence, after compression, we perform benchmarks on three systems, which were previously studied: the Mo-Nb-Ta-W medium-entropy alloy \cite{hodapp2024-ETN}, the molten salt mixture LiF-NaF-KF~\cite{rybin2024-FLiNaK}, and glycine~\cite{rybin2025-mol-cryst}. These systems were chosen not only because the datasets are available, but because MLIPs are, in principle, frequently used to model properties of alloys~\cite{rosenbrock2021machine, kostiuchenko2019impact},  molten salts~\cite{porter2022computational}, and molecular crystals. In the former case, particularly to accelerate the search for the most energetically favorable polymorphs~\cite{kapil2022complete, rybin2025-mol-cryst, D5SC01325A, D5SC00677E}. The success of such calculations is highly dependent on the correct energy ranking of the polymorphs~\cite{hunnisett2024seventh}. Molecular crystals typically exhibit complex polymorphic energy landscapes, with numerous structures located within an energy window of a few kilojoules per mole~\cite{reilly2016report}. Consequently, a model should have DFT accuracy to reliably range polymorphs, which is a good test for compressed MLIP. 

We mainly test our methods using Moment Tensor Potential (MTP) as an example model, which is frequently used in computational studies (see, e.g., Refs.\cite{hodapp2024-ETN,rybin2024-FLiNaK,rybin2025-mol-cryst}). This MLIP has two types of parameters to fit: linear and radial. The radial parameters are tensors of the fourth order, and we take advantage of tensor-based decompositions to compress these radial parameters. We demonstrate that the algorithms used enable both reduction of the number of MTP parameters and improvement of accuracy of MTP fitting for the three systems considered. Without loss of generality, the established methodology can be used to compress other MLIPs, for example, ACE~\cite{drautz2019-ACE}. In addition to MTP, we apply the methodology to ACE and on the example of the FLiNaK system we demonstrate that the compressed ACE retains the accuracy of the base ACE.

\section{Methodology}

\subsection{Notations}

This subsection defines the basic mathematical notation and operations used in this work.

General rank matrix factorization of $A \in \mathbb{R}^{m \times n}$, also known as the skeleton decomposition, is read as 
\[A = BC, \quad B \in \mathbb{R}^{m \times k}, \quad C \in \mathbb{R}^{k \times n},\]
where $k \leq \min(m,n)$ is the rank of $A$. 
This factorization is not unique, and additional constraints, such as the orthogonality of the columns in $B$ and $C$ may be imposed.  
As a result, the skeleton decomposition can always be reduced to the form of a singular value decomposition (SVD), defined as: \[A = U\Sigma V^\top,\] where $U \in \mathbb{R}^{m \times k}$ and $V \in \mathbb{R}^{n \times k}$ are matrices with orthonormal columns, and $\Sigma \in \mathbb{R}^{k \times k}$ is a diagonal matrix with singular values $\sigma_1 \geq \sigma_2 \geq \cdots \geq \sigma_k > 0$ on the diagonal. 

The inner product of two matrices is $\langle A, B \rangle = \operatorname{tr}(A^\top B)$ and the corresponding induced norm, known as the Frobenius norm, has the following form: \[\|A\|_F = \sqrt{\langle A, A \rangle}.\]
The best rank-$r$ approximation of a matrix $A$ with respect to the Frobenius norm is given by the so-called truncated SVD: 
\[A_r = U_r\Sigma_r V_r^\top,\] where $U_r$ and $V_r$ contain the first $r$ columns of $U$ and $V$, respectively, and $\Sigma_r$ is the leading principal $r\times r$ submatrix of~$\Sigma$. By $U_r^\perp$, $V_r^\perp$\label{not:orth_matr_perp} we denote matrices that contain all the columns of $U$, $V$ besides the leading $r$. In other words, 
\begin{equation}\label{not:orth_matr_perp}
    U = [\underbrace{U_r}_{r}\ \underbrace{U_r^\perp}_{m-r}], \quad V = [\underbrace{V_r}_{r}\ \underbrace{V_r^\perp}_{n-r}].
\end{equation} 
We also use the notation 
\[A_r = \mathrm{SVD}_r(A).\]

By a $d$-dimensional tensor, we imply a multidimensional array:
\[A = \{a_{i_1\ldots i_d}\}_{i_1,\dots,i_d=1}^{n_1,\dots,n_d}\in \mathbb{R}^{n_1 \times \ldots \times n_d}.
\] 
We denote the outer (tensor) product of two tensors by ``$\circ$''. An outer product of $u_1 \in \mathbb{R}^{n_1}, \ldots, u_d \in \mathbb{R}^{n_d}$ is 
\[
A = u_1 \circ \ldots \circ u_d \in \mathbb{R}^{n_1 \times \ldots \times n_d}
\] 
with the entries
\[
a_{i_1\ldots i_d} = (u_1)_{i_1}\cdot \ldots \cdot (u_d)_{i_d}.
\] 
For $k$ equal multipliers, we use the notation \[u^{\circ k} = u \circ \dots \circ u.\] 

The reshape operation of $A \in \mathbb{R}^{n_1 \times \cdots \times n_d}$ has the following form:
\[
\mathrm{reshape}({A}, [m_1, \dots, m_k]) = {B} \in \mathbb{R}^{m_1 \times \cdots \times m_k},
\]
where \( \prod_{i=1}^d n_i = \prod_{j=1}^k m_j \), reorders the elements of \( {A} \) into \( {B} \) such that the flattened underlying array remains identical under lexicographical (row-major) ordering of indices.

Finally, by orthogonal projector $P$ we imply a matrix, satisfying $P^2 = P$ and $P^\top = P$.

\subsection{Moment tensor potential}

In this subsection we describe a machine-learning interatomic potential (MLIP) compressed in the study. Moment Tensor Potential (MTP) is a MLIP that was originally proposed for single-component materials\cite{shapeev2016-mtp} and generalized to multi-component materials\cite{gubaev2018-multMTP}.

Energy of the system in the case of MTP usage $E^{\rm MTP}$ is the sum of contributions $V^{\rm MTP}(\mathfrak{\bm n}_i)$ of atomic neighborhoods ${\bf \mathfrak{n}}_i$ for $N$ atoms
\begin{align} \label{EnergyMTP}
E^{\rm MTP} = \sum \limits_{i=1}^{N} V^{\rm MTP}(\mathfrak{\bm n}_i). 
\end{align}
Each neighborhood is a tuple 
$$\mathfrak{ n}_i = ( \{r_{i1},z_i,z_1 \}, \ldots, \{r_{ij},z_i,z_j \}, \ldots, \{r_{iN_ {\rm nbh}},z_i,z_{N_ {\rm nbh}} \} ),$$ 
where $r_{ij}$ are relative atomic positions, $z_i$, $z_j$ are the types of central and neighboring atoms, and $N_ {\rm nbh}$ is the number of atoms in neighborhood. We denote the number of atomic types in a system by $N_T$. Each contribution $V^{\rm MTP}(\mathfrak{\bm n}_i)$ in the potential energy $E^{\rm MTP}$ expands through a set of basis functions
\begin{align} \label{SiteEnergyMTP}
V^{\rm MTP}({\bf \mathfrak{n}}_i) = \sum \limits_{\alpha} \xi_{\alpha} B_{\alpha}({\mathfrak{\bm n}}_i),
\end{align} 
where $B_{\alpha}$ are the MTP basis functions, $\Xi = \{ \xi_{\alpha} \}$ are the linear parameters to be found. To define the functional form of the MTP basis functions we introduce the so-called moment tensor descriptors:
\begin{equation}\label{MomentTesnsorDescriptors}
M_{\mu,\nu}({\mathfrak{\bm n}}_i)=\sum_{j=1}^{N_{\rm nbh}} f_{\mu}(|r_{ij}|,z_i,z_j) r_{ij}^{\circ \nu}.
\end{equation}
The descriptor consists of the angular part $r_{ij}^{\circ \nu}$, which is the tensor of $\nu$-th order, and the radial part $f_{\mu}(|r_{ij}|,z_i,z_j)$ of the following form:
\begin{align} \label{RadialFunction}
\displaystyle
f_{\mu}(|r_{ij}|,z_i,z_j) = \sum_{\beta=1}^{N_b} c^{(\beta)}_{\mu, z_i, z_j} T^{(\beta)} (|r_{ij}|) (R_{\rm cut} - |r_{ij}|)^2.
\end{align}
Here $\mu$ is the number of the radial function $f_{\mu}$, $C=\{c^{(\beta)}_{\mu, z_i, z_j}\}$ are the radial parameters to be found, $T^{(\beta)} (|r_{ij}|)$ are polynomial functions $\beta=1,\ldots,N_b$ (where $N_b$ is the number of polynomial functions), and the term $(R_{\rm cut} - |r_{ij}|)^2$ is introduced to ensure smoothness with respect to the atoms leaving and entering the sphere with the cutoff radius $R_{\rm cut}$. We denote the number of radial functions by $N_f$.
 
By definition, the MTP basis function $B_{\alpha}$ is a contraction of one or more moment tensor descriptors, yielding a scalar. To construct the basis functions $B_{\alpha}$ and determine a particular functional form of MTP, we define the so-called level of the moment tensor descriptor:
\begin{equation} \label{eq:LevelMTD}
\displaystyle
{\rm lev} M_{\mu,\nu} = 2 + 4 \mu + \nu.
\end{equation}
We also define the level of the MTP basis function:
\begin{equation} \label{LevelMultMTD}
\displaystyle
{\rm lev} B_{\alpha} = \rm {lev} \underbrace {\prod_{p=1}^{P} M_{\mu_p,\nu_p}}_{scalar} = \sum \limits_{p=1}^P (2 + 4 \mu_p + \nu_p).
\end{equation}
A set of MTP basis functions and, thus, a particular functional form of MTP depends on the maximum level, ${\rm lev_{\rm max}}$, which we also call the level of MTP. In the set of MTP basis functions, we include only those with ${\rm lev} B_{\alpha} \leq {\rm lev_{\rm max}}$. The level of MTP determines the number of linear parameters $\Xi$ and the number of radial functions $N_f$ as the moment tensor descriptors depend on the number of the radial function. The set of MTP parameters to be found is denoted by ${\boldsymbol \Theta} = (\Xi, C)$ and the MTP energy of a structure is denoted by $E^{\text{MTP}} = E({\boldsymbol \Theta}) = E(\Xi, C) $. We further refer to this MTP as the base MTP.

A drawback of the functional form of the base MTP is the radial parameters $c^{(\beta)}_{\mu, z_i, z_j}$. The number of these parameters is $N_T^2 N_f N_b$, i.e., it scales quadratically with the number of atomic types $N_T$ in a system. In this work, we demonstrate that a certain percentage of these parameters is redundant and that the same or even smaller training errors can be achieved with the reduced number of them. To reduce the number of radial parameters we use two techniques. The first one is based on the so-called low-rank matrix and tensor approximations yielding a low-rank MTP. The second technique is the so-called Riemannian optimization. We describe them in the following subsections.

\subsection{Matrix and tensor factorization of radial parameters}

Here we describe two ways for reducing the number of the radial parameters $C\;\in\;\mathbb{R}^{N_T\times N_T\times N_f\times N_b}$. These approaches are matrix factorization in the form of the skeleton decomposition of the reshaped radial parameters and factorization of the tensor of the radial parameters in the form of the matrix product state (MPS)~\cite{schollwock2011density}, which is also known under the name tensor train (TT) decomposition~\cite{oseledets2011tensor} in applied linear algebra.
Notably, low-rank tensor factorizations have also been used in the context of Hartree-Fock and DFT calculations for compressing coefficient tensors~\cite{khoromskaia2015tensor,khoromskij2011numerical,rakhuba2016grid}.

Matrix factorization (MF) approximates a given matrix $A \in \mathbb{R}^{m \times n}$ by the product of two smaller matrices:
$$ A \approx UV^\top, $$
where $U \in \mathbb{R}^{m \times r}$ and $V \in \mathbb{R}^{n \times r}$, and $r$ denotes the chosen rank.

To use a matrix factorization, the four-dimensional tensor $c^{(\beta)}_{\mu, z_i, z_j}$ is first ``unfolded'' into a matrix
$$
\widehat{C} \;=\;\mathrm{reshape}\bigl(C,\;[N_T\cdot N_T, N_f\cdot N_b]\bigr)
\;\in\;\mathbb{R}^{(N_T^2)\times (N_fN_b)},
$$
by merging the first two indices into one composite index of length $N_T^2$ and the last two into one of length $N_fN_b$. We then approximate the matrix of radial parameters $\widehat C \;\approx\; U\,V^{\!\top}$. 
The base MTP contains $N_T^2 N_f N_b$ radial parameters, and its factorized representation stores only $r(N_T^2 + N_f N_b)$ radial parameters. 

Tensor factorization (TF) approach is based on MPS/TT-decomposition. TT format expresses $C$ as a contraction of four third-order core tensors:
\[
G^{(k)} \;\in\; \mathbb{R}^{r_{k-1}\times n_k \times r_k}
\quad
\text{for }k=1,\ldots,4,
\quad
r_0 = r_4 = 1,
\]
so that:
\[
c^{(\beta)}_{z_i,z_j,\mu}
=
\sum_{\alpha_1=1}^{r_1}
\sum_{\alpha_2=1}^{r_2}
\sum_{\alpha_3=1}^{r_3}
G^{(1)}_{\,1,\,z_i,\,\alpha_1}
\;G^{(2)}_{\alpha_1,\,z_j,\,\alpha_2}
\;G^{(3)}_{\alpha_2,\,\mu,\,\alpha_3}
\;G^{(4)}_{\alpha_3,\,\beta,\,1}.
\]

The total number of the radial parameters in the TT format is given by:
\[
\sum_{k=1}^{4} r_{k-1}n_kr_k \;=\; 
r_0 N_T r_1 +
r_1 N_T r_2 +
r_2 N_f r_3 +
r_3 N_b r_4, r_0=r_4=1,
\]
i.e., if $r_1=r_2=r_3=r$ then we have $(N_T+N_f) r^2 + (N_T+N_b)r$ radial parameters instead of $N_T^2 N_f N_b$ in the base MTP. The choice of TT‐ranks \(\{r_1,r_2,r_3\}\) governs the trade‐off between representational flexibility and compactness. 

\subsection{Fitting the potential}

We begin this subsection by formalizing the loss function. Subsequently, we present a discussion of the optimization techniques employed to minimize this function subject to low-rank constraints, considering two distinct scenarios.
We first describe optimization under a fixed-rank constraint, followed by a method to gradually increase the rank.

\subsubsection{Loss function}

To find optimal parameters ${\bar{{\bm {\Theta}}}}$ of MTP, we solve the following optimization problem (minimization of the loss function):
\begin{equation} \label{Fitting}
\begin{array}{c}
\displaystyle
L({\bm {\Theta}}) = \sum \limits_{k=1}^K \Bigl[ w_{\rm e} \left(E_k({\bm {\Theta}}) - E^{\rm DFT}_k \right)^2 +
\\
\displaystyle
w_{\rm f} \sum_{i=1}^{N_k} \left| {\bf f}_{i,k}({\bm {\Theta}}) - {\bf f}^{\rm DFT}_{i,k} \right|^2 +
\\
\displaystyle
w_{\rm s} \sum_{i=1}^6 \left| \sigma_{i,k}({\bm {\Theta}}) - \sigma^{\rm DFT}_{i,k} \right|^2
\Bigr]  \to \operatorname{min},
\end{array}
\end{equation}
where $K$ is a number of configurations in the training set and $N_k$, $k=1,\ldots,K$ is a number of atoms for each configuration, $E^{\rm DFT}_k$, ${\bf f}^{\rm DFT}_{i,k}$, and $\sigma^{\rm DFT}_{i,k}$ are the DFT energies, forces, and stresses to which we fit the MTP energies $E_k({\bm {\Theta}})$, forces ${\bf f}_{i,k}({\bm {\Theta}})$, and stresses $\sigma_{i,k}({\bm {\Theta}})$, thus optimizing the MTP parameters ${\bm {\Theta}}$. The factors $w_{\rm e}$, $w_{\rm f}$, and $w_{\rm s}$ in \eqref{Fitting} are non-negative weights which express the importance of energies, forces, and stresses with respect to each other.

We find the optimal parameters ${\bar{{\bm {\Theta}}}}$ numerically, using the iterative method to minimize the non-linear loss function, as is described further in this section. 

\subsubsection{Optimization with a fixed-rank constraint}

\paragraph{MF optimization.} 
When the rank is fixed, our goal is to minimize the loss function with respect to the factor matrices $U$ and $V$ with $r$ columns:
\[
     {L}\big(\bm{\Theta}) = 
     {L}\big(\Xi, \widehat C \big) = 
     {L}\big(\Xi, UV^\top\big) \equiv  
     {L}\big(\Xi, U, V\big) \to \min_{\Xi, U, V}.
\]
The derivatives of the loss function are calculated with respect to these parameters $\Xi, U, V$ as follows:
$$
\frac{\partial L}{\partial \Xi} = \frac{\partial L}{\partial \Xi},
\quad \frac{\partial L}{\partial U}
=\frac{\partial L}{\partial \widehat C}\,V,
\quad \frac{\partial L}{\partial V}
=\Bigl(\frac{\partial L}{\partial \widehat C}\Bigr)^{\!\top}U.
$$
As it could be seen, we use the original derivative ${\partial L}/{\partial \widehat C}$ to calculate the new derivatives ${\partial L}/{\partial U}$ and ${\partial L}/{\partial V}$.
We further refer to MTP based on the described matrix factorization as the MF MTP or simply MF for short.

\paragraph{R-MF optimization.} 
Let us discuss the so-called Riemannian version of matrix factorization, which we abbreviate as R-MF. 
It is important to note that the parameters $(U, V)$ are not unique. This is because many different pairs of matrices $(U, V)$ can produce the same product $UV^\top$. For example, for any invertible $r \times r$ matrix $S$, we can define new matrices:
\[
\widetilde U = US^{-1}, \quad \widetilde V^\top = S V^\top,
\]
and the product remains unchanged: $\widetilde U \widetilde V^\top = UV^\top$. This means that the same solution can be represented in an infinite number of ways, which can affect the optimization problem.
To avoid this issue, we can optimize directly over the set of all fixed-rank matrices:
\[
\mathcal{M}_r = \{ X \in \mathbb{R}^{m \times n} : \mathrm{rank}(X) = r \}.
\]
The set $\mathcal{M}_r$ constitutes a smooth manifold. Its local linearization at a point $X \in \mathcal{M}_r$ is the tangent space $T_X \mathcal{M}_r$. By equipping each tangent space with a natural inner product $\langle \cdot, \cdot \rangle$ the manifold $\mathcal{M}_r$ becomes a Riemannian manifold. Riemannian optimization leverages this geometry to find the minimum of a function on $\mathcal{M}_r$, effectively ignoring the unnecessary degrees of freedom in the factors $U$ and $V$.

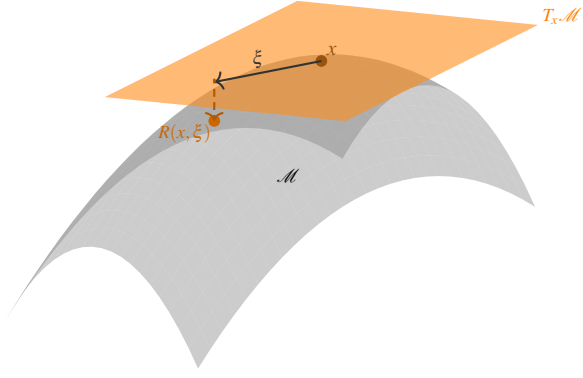
\begin{figure}[t]
    \centering
    \input{retraction.tikz}
    \caption{Illustration of the concepts of a tangent space $T_x \mathcal{M}$ and a retraction $R(x,\xi)$ for a smooth manifold $\mathcal{M}$.}
    \label{fig:retraction}
\end{figure}

While classical optimization involves moving in a straight line within a vector space, optimization on a Riemannian manifold $\mathcal{M}_r$ requires moving along the curved geometry of the constraint set. The process is generalized as follows: first, a search direction $\xi$ is chosen within the so-called tangent space $T_X\mathcal{M}_r$ (local linearization of $\mathcal{M}_r$) at the current point $X$. Since moving directly along this tangent vector would leave the manifold, a retraction operator $R_X$ is used to map the tangent vector $\xi$ onto $\mathcal{M}_r$:
$$
R_{X}(\xi) = \operatorname{SVD}_r(X + \xi),
$$
see also Ref.~\cite{absil2015low} for alternative retractions. 
The illustration of the concept is shown on Figure~\ref{fig:retraction}.

We propose to compute the search direction $\xi$ using a Riemannian Broyden–Fletcher–Goldfarb–Shanno (RBFGS) method~\cite{qi2010riemannian}, adapted for matrix manifolds. A complete description of the algorithm is provided in the supplementary material. 
 For more details regarding the Riemannian optimization on matrix manifolds, see Ref. \cite{absil2009optimization}.

\paragraph{TF optimization.}
The tensor case is similar to MF, except that we minimize the loss function $L(\Xi,G^{(1)},G^{(2)},G^{(3)},G^{(4)})$ with respect to all TT‐cores $G^{(k)}, k=1,\ldots,4$ and linear parameters $\Xi$ using the BFGS algorithm. Gradients of the loss function with respect to each core are given in the supplementary material. We also note that for consistency with matrix factorization (MF) and to improve clarity in the tables and figures, we use the notation TF (tensor factorization) instead of TT.

We denote MTP in which the tensor of radial parameters is represented in TT-format by the TF MTP. The Riemannian version of TF is also possible, but we found that it was more computationally demanding than R-MF, while not giving addition boosts in terms of quality of the optimal solution. Therefore, we fully omit this case in our work.

\subsubsection{Optimization with rank augmentation} 

We observed that by gradually increasing the rank value, we are capable of finding better optima. Here, we describe an algorithm. 
Denote \(r_{\min},r_{\max},s,\Delta r\) by initial rank, maximum rank, augmentation interval during BFGS iterations, or the number of BFGS steps conducted with the current rank, and rank increment, respectively. We randomly initialize
\[
U\;\in\;\mathbb{R}^{(N_T^2)\times r_{\min}},\quad
V\;\in\;\mathbb{R}^{(N_fN_b)\times r_{\min}},
\]
and start from $r=r_{\min}$. After each of the $s$ iterations of the BFGS algorithm applied to minimize the loss function $L(\Xi, U, V)$, we perform the adaptive update steps if $r < r_{\max}$, where we enrich our approximation with a gradient information. The augmented update steps sequentially increase the rank of the model. A similar approach was used, e.g., in Ref.\cite{dolgov2014alternating} for multidimensional linear systems. These steps are presented in detail in Algorithm~\ref{alg:inc-rank}. At each step, the model stores only $r (N_T^2 + N_fN_b)$ radial parameters in factors $U$ and $V$. We refer to the MTP obtained with Algorithm ~\ref{alg:inc-rank} as the MFRA MTP.

\begin{algorithm}[ht]
\caption{MFRA}
\label{alg:inc-rank}
\KwIn{Tensor $ C \in \mathbb{R}^{N_T\times N_T\times N_f\times N_b}$; 
initial rank $r_{\min}$; maximum rank $r_{\max}$; augmentation interval $s$; 
rank increment $\Delta r$; loss $L(\Xi, \widehat C)$.}
\KwIn{$r \gets r_{\min}$; 
$U\in\mathbb{R}^{(N_T^2)\times r},\;
V\in\mathbb{R}^{(N_fN_b)\times r}$ initialized randomly.}

\For{$k = 1,2,\dots$}{
    Take one BFGS update on 
    $L(\Xi, U, V) = L(\Xi, UV^{T})$ to refine $U,V$\;
    
    \If{$k \bmod s = 0$ \textbf{and} $r < r_{\max}$}{
        Form approximation $\widehat C \gets U V^\top$\;
        
        Compute gradient 
        $G \gets \dfrac{\partial L}{\partial \widehat C}\,\in\,\mathbb{R}^{(N_T^2)\times(N_fN_b)}$\;
        
        Compute SVD 
        $G \approx \widetilde U\,\Sigma\,\widetilde V^\top$\;
        
        Extract top-$\Delta r$ singular vectors:
        $U_{\Delta} \gets \widetilde U[:,:\Delta r],\;
         V_{\Delta} \gets \widetilde V[:,:\Delta r],\;
         \Sigma_{\Delta} \gets \Sigma[:\Delta r, :\Delta r]$\;
        
        Use line search to find
        $h = \arg\min_{h'}\,L\bigl(\Xi, UV^\top + h'\,U_{\Delta} \Sigma_{\Delta} V_{\Delta}^\top\bigr)$\;
        
        Increase rank:
        $U \gets [\,U \;\big|\; h U_{\Delta} \Sigma_{\Delta}^{1/2}],\;
         V \gets [\,V \;\big|\; V_{\Delta} \Sigma_{\Delta}^{1/2}],\;
         r \gets r + \Delta r$\;
    }
}
\end{algorithm}

\subsection{Compressed MTP and Equivariant Tensor Network}

In Ref.\cite{hodapp2024-ETN}, Equivariant Tensor Network (ETN) potential was proposed. An energy $V^{\rm ETN}_i \equiv V^{\rm ETN}({\bf \mathfrak{n}}_i)$ of atomic neighborhood ${\bf \mathfrak{n}}_i$ in ETN potential is the following tensor train (TT):
\begin{multline}\label{eq:etn_pot}
    V^{\rm ETN}_i
    =
    \sum_{k_1} \sum_{k_2} \ldots \sum_{k_{d-2}} \sum_{k_{d-1}} \left(\sum_{k_1'} T^1_{k_1' k_1} v_{k_1'}\right) 
    \left(\sum_{k_2'} T^2_{k_1 k_2' k_2} v_{k_2'}\right)
    \ldots \\
    \displaystyle
    \left(\sum_{k_{d-1}'} T^{d-1}_{k_{d-2} k_{d-1}' k_{d-1}} v_{k_{d-1}'}\right) \left(\sum_{k_d'} T^d_{k_{d-1} k_d'} v_{k_d'}\right),
\end{multline}
where $T^1_{k_1' k_1}, T^2_{k_1 k_2' k_2}, \ldots, T^{d-1}_{k_{d-2} k_{d-1}' k_{d-1}}, T^d_{k_{d-1} k_d'}$ are the TT cores, and $v_{k_j'} \equiv v_{k_j'}({\bf \mathfrak{n}}_i)$, $j=1,\ldots,d$ are the following feature vectors:
\begin{equation}\label{eq:etn_feature}
    v_{k} \equiv v_{lmn}
    =
    \sum_j f_n(|r_{ij}|,z_i,z_j) Y_{lm}(r_{ij} / |r_{ij}|),
\end{equation}
where a multi-index $k=(lmn)$ includes the index $l = 0, \ldots, L$ of the subspace of the irreducible representation, the dimension $m = -l, -l + 1, \ldots, l$ of the subspace, and the number $n = 1, \ldots, N(l)$ of the radial channels corresponding to each $l$ (we denote the maximum number of the radial channels by $N_{\rm rad}^{\rm max}$), $f_n(|r_{ij}|,z_i,z_j)$ is a radial part describing the pairwise interactions, and $Y_{l m}(r_{ij} / |r_{ij}|)$ are spherical harmonics describing an angular part. The $n$-th radial function in the ETN feature vector has the form:
\begin{equation}\label{eq:etn_radial_basis}
    f_n(r_{ij},z_i,z_j) = \sum_{\beta\lambda} B_{n \beta \lambda} \varphi^{(\beta)}(|r_{ij}|) A_{\lambda z_i z_j}.
\end{equation}
Here, $\varphi^{(\beta)}(|r_{ij}|)$ are polynomial functions, $B_{n \beta \lambda}$ and $A_{\lambda z_i z_j}$ are sparse tensors to be fitted (see details in Ref. \cite{hodapp2024-ETN}). Thus, in the ETN potential the coefficients in the radial part are the compressed tensors by its construction whereas in the compressed MTP the original tensor of the radial parameters is compressed using the known low-rank approximations.

A sufficient condition to make $V^{\rm ETN}_i$ invariant to rotations is to let the TT cores be equivariant mappings of the feature vectors to rotate them correspondingly with a basis change under SO(3). To that end, the Wigner-Eckhart theorem is employed. This theorem states that every tensor with three multi-indices $\{ (l_i, m_i, n_i) \}_{i=1,2,3}$ can be represented in the form of equivariant mapping:
\begin{equation}\label{eq:etn_core}
    T_{k_1 k_2 k_3}
    =
    \xi_{(l_1 n_1)(l_2 n_2)(l_3 n_3)}
    C_{(l_1 m_1)(l_2 m_2)(l_3 m_3)}
    ,
\end{equation}
where $\xi_{(l_1 n_1)(l_2 n_2)(l_3 n_3)}$ is the tensor of coefficients to be fitted, and $C_{(l_1 m_1)(l_2 m_2)(l_3 m_3)}$ is the Clebsch-Gordan coefficient that defines the symmetry group.

To improve accuracy of ETN, we can increase a number of TT cores in \eqref{eq:etn_pot} and a number of parameters in the tensors of the radial coefficients in \eqref{eq:etn_radial_basis}. To obtain an MTP of a reasonable accuracy, we can increase its level and add more basis functions $B_{\alpha}$ in its functional form. This leads to an uncontrolled growth of the number of linear and radial parameters in the base MTP. However, in the compressed MTPs, it is possible to control a number of radial parameters by applying low-rank approximations to the tensor of the radial parameters. Thus, in this work we developed the methods to control the growth of the number of parameters in MTP. We illustrate this fact in the section with the results below.

\subsection{Compressed Atomic Cluster Expansion}

To demonstrate the applicability of the proposed methodology for compression of other MLIPs, we use it for the compression of the radial part of Atomic Cluster Expansion (ACE) from Ref. \cite{drautz2019-ACE}. To construct ACE, a single bond basis function is introduced: 
\begin{equation}\label{ACE_phi}
\phi_{z_i z_j nlm}(\bm{r}_{ij}) = R_{nl}^{z_i z_j} (|\bm{r}_{ij}|) Y_{lm} (\bm{r}_{ij}/|\bm{r}_{ij}|),
\end{equation}
where $R_{nl}^{z_i z_j} (|\bm{r}_{ij}|)$ is a radial part and $Y_{lm} (\bm{r}_{ij}/|\bm{r}_{ij}|)$ are spherical harmonics describing an angular part. The meaning of the indices $l$, $m$, $n$ here is the same as in ETN described above. The single bond atomic basis function has the following form:
\begin{equation}\label{ACE_basis_single}
A_{iz nlm} = \sum \limits_{j=1}^{N_{\rm nbh}} \delta_{z z_j} \phi_{z_i z_j nlm}(\bm{r}_{ij}).
\end{equation}
Permutation-invariant many-body basis functions are generated by taking products as follows:
\begin{equation}\label{ACE_basis}
\bm{A}_{i \bm{z} \bm{n} \bm{l} \bm{m}} = \prod_{t=1}^{\nu} A_{i z^t n^t l^t m^t }.
\end{equation}
The products follow a body order of $\nu+1$. The vectors $z$, $n$, $l$, $m$ have a length of $\nu$. Finally, the energy $E_i$ of atom $i$ is:
\begin{equation}\label{ACE_site_energy}
E_i = \sum \limits_{ \bm{z} \bm{n} \bm{l} \bm{m} } \tilde{\bm{c}}_{z_i \bm{z} \bm{n} \bm{l} \bm{m}} \bm{A}_{i \bm{z} \bm{n} \bm{l} \bm{m}},
\end{equation}
where $\tilde{\bm{c}}_{z_i \bm{z} \bm{n} \bm{l} \bm{m} }$ are the linear parameters of ACE to be fitted. In this work, we refer to the above ACE as the base ACE.

As for MTP, the radial part of ACE is also an expansion over the polynomial functions $\phi_k$:
\begin{equation}\label{ACE_site_energy}
R_{nl}^{z_i z_j} (|\bm{r}_{ij}|) = \sum_{k=1}^{N_k} c_{nlk}^{z_i z_j} \phi_k(|\bm{r}_{ij}|),
\end{equation}
where $c_{nlk}^{z_i z_j}$ are the radial parameters to be fitted, $N_k$ is the number of polynomials. The total number of these parameters is $(N_T(N_T-1)/2 + N_T) N_k N_{\rm rad}^{\rm max} (L+1)$ and this number consists of $(N_T(N_T-1)/2 + N_T)$ tensors $C^{\rm ACE}$ of the third order including $N_k N_{\rm rad}^{\rm max} (L+1)$ parameters. We compressed these tensors of the third order using the matrix factorization. To that end, we reshaped them into a matrix 
$$\hat{C}^{\rm ACE} = \mathrm{reshape}\bigl(C^{\rm ACE},\;[N_k, N_{\rm rad}^{\rm max} \cdot (L+1)]\bigr)
\;\in\;\mathbb{R}^{N_k\times N_{\rm rad}^{\rm max}(L+1)}.$$
Similarly to the MF MTP, we then approximate the matrix of the radial parameters $\widehat C^{\rm ACE} \;\approx\; U\,V^{\!\top}$ where $U \in \mathbb{R}^{N_k\times r}$, $V \in \mathbb{R}^{N_{\rm rad}^{\rm max}(L+1)\times r}$, and $r$ is the chosen rank. 
The compressed ACE contains only $r(N_k + N_{\rm rad}^{\rm max}(L+1))$ radial parameters. To optimize this potential, we use the same algorithm as for the MF MTP.

\section{Results and discussion}

\subsection{Computational details}

We applied developed methods to three systems: the Mo-Nb-Ta-W medium-entropy alloy, the molten salt LiF-NaF-KF (FLiNaK) mixture, and different polymorphs of glycine. The training sets were taken from Refs.~\cite{hodapp2024-ETN, rybin2024-FLiNaK, rybin2025-mol-cryst}. These data sets were calculated using the VASP package \cite{VASP} with the projector-augmented wave method~\cite{Kresse1999}. The Perdew-Burke-Ernzerhof generalized gradient approximation (PBE)~\cite{Perdew1996} was used as the exchange–correlation functional, and the DFT-D3 method~\cite{Grimme2010} was utilized to account for the dispersion forces in the case of FLiNaK and glycine modeling. A brief description of the calculation parameters is given in Tabel~\ref{tab:dataset_generation} and details might be found in the aforementioned works. We also note that for the Mo-Nb-Ta-W system we have a validation set of 546 configurations.

\begin{table}[!ht]
    \caption{Computational details on the generation of the training sets.}
    \begin{center}
    % \centering
    \begin{tabular}{c|c|c|c|c}
    \hline
    \hline
    System & Level & Cutoff, & K-points grid & Training set \\
     & of theory & eV & & size
    \\
    \hline
    Mo-Nb-Ta-W & PBE & 520 & 4 $\times$ 4 $\times$ 4 & 4983 
    \\
    FLiNaK & PBE+D3 & 550 & $\Gamma$-point & 754 
    \\
    glycine & PBE+D3 & 600 & $\Gamma$-point & 3127 
    \\
    \hline
    \hline
    \end{tabular}
    \end{center}
    \label{tab:dataset_generation}
\end{table}

%To demonstrate how to choose the optimal ranks of different MTPs, 
We trained an ensemble of five MTPs of each type, namely, base, MF, R-MF, MFRA, and TF to investigate Mo-Nb-Ta-W and FLiNaK, and we trained ensembles of 10 MTPs for the glycine system. We fixed the rank $r$ for the calculations with MF, R-MF, and TF. For all calculations with MFRA, we took $s=80$ and $r_{min}=\Delta r$ and increased the rank of the potential until we reached a predetermined $r=r_{max}$. We continued further calculations with this rank. Practical recommendations for the choice of the hyperparameters of MFRA are given in the supplementary material. We took $N_b = 8$ for all MTPs. The rest of details on choosing the optimal ranks are given in the next subsection. 

For the Mo-Nb-Ta-W system we also trained ensembles of five ETN potentials and compared the dependence of validation errors in predicting energy and forces on the number of parameters for ETN and compressed MTPs. For the FLiNaK system we fitted the base ACE and its compressed version and demonstrated that the matrix factorization can be applied to compression of the radial parameters of ACE. We took ACE with $N_k=9$, $N_{\rm rad}^{\rm max}=3$, and $L=2$. All the potentials mentioned in this study were fitted with $R_{\rm cut}=5$ \AA.

We used the weights $w_e=1$ eV$^{-2}$, $w_f=0.01$ (eV/\AA)$^{-2}$, and $w_s=0$ eV$^{-2}$ for the fitting on the Mo-Nb-Ta-W and FLiNaK training sets, in the case of glycine we also took into account stresses with $w_s=0.001$ eV$^{-2}$, since in the latter case we had to perform structural relaxation as demonstrated below. We utilized the BFGS algorithm for fitting of all the models, the number of BFGS steps was 2000.

\begin{figure*}[!ht]
	\centering
	\begin{minipage}[h]{0.33\linewidth}
		\center{\includegraphics[trim={0cm 0cm 0cm 0cm},clip, width=1\linewidth]{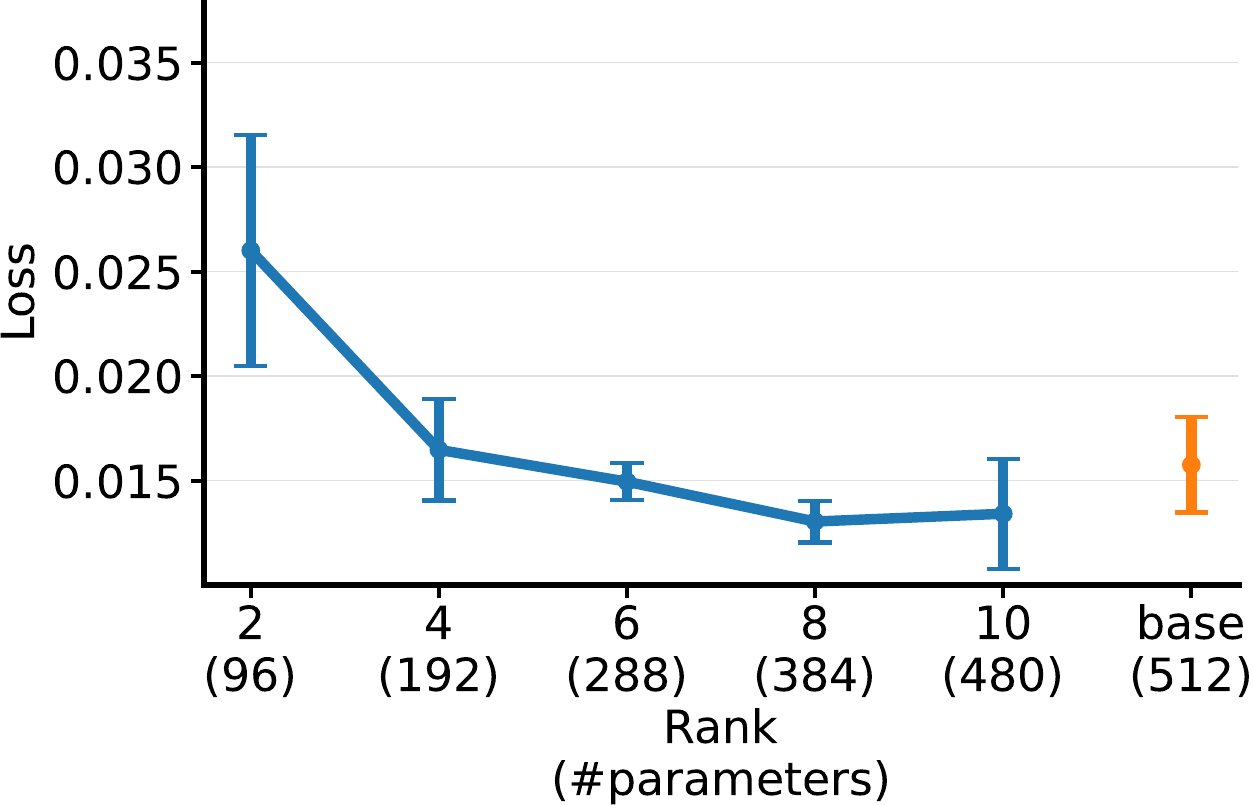}} \\ (a)
	\end{minipage}
    \begin{minipage}[h]{0.33\linewidth}
		\center{\includegraphics[trim={0cm 0cm 0cm 0cm},clip, width=1\linewidth]{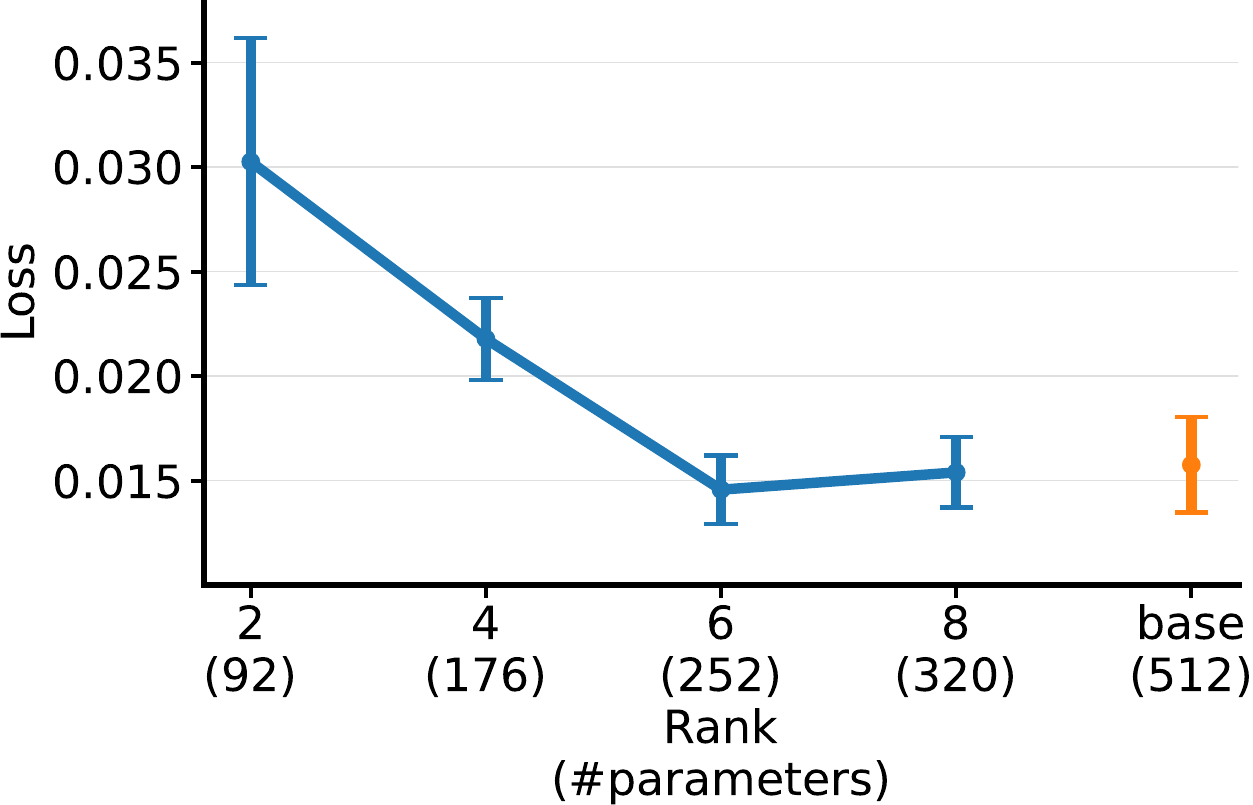}} \\ (b) 
	\end{minipage}
    \begin{minipage}[h]{0.33\linewidth}
		\center{\includegraphics[trim={0cm 0cm 0cm 0cm},clip, width=1\linewidth]{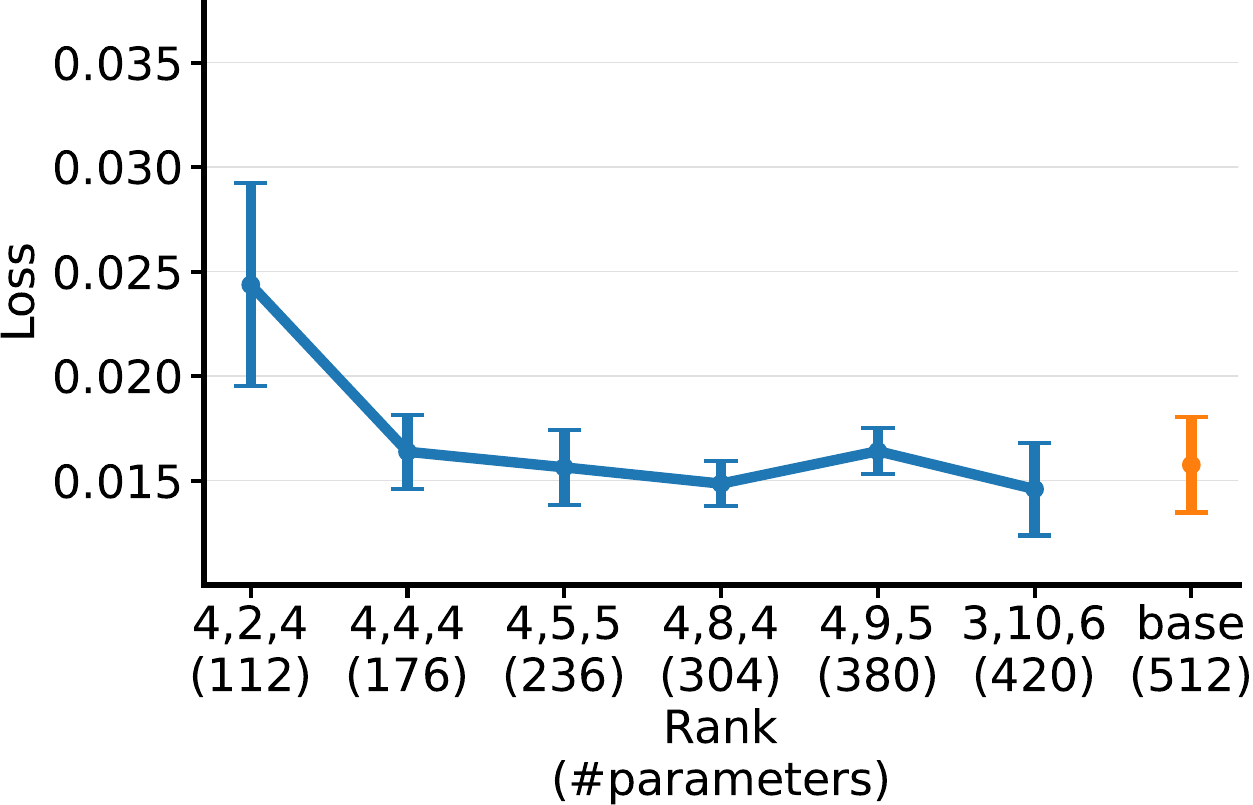}} \\ (c)
	\end{minipage}
    \caption[]{Dependence of the loss function calculated on the Mo-Nb-Ta-W validation set on the rank of (a) the MF (b) the R-MF (c) the TF MTPs of the 14th level. The loss function for the base MTP is also shown. Error bars demonstrate uncertainty of the loss function prediction and are given within 1-$\sigma$ confidence interval.} 
    \label{fig:optimal_rank}
\end{figure*}

\begin{figure*}[!ht]
    \centering
    \includegraphics[width=\textwidth]{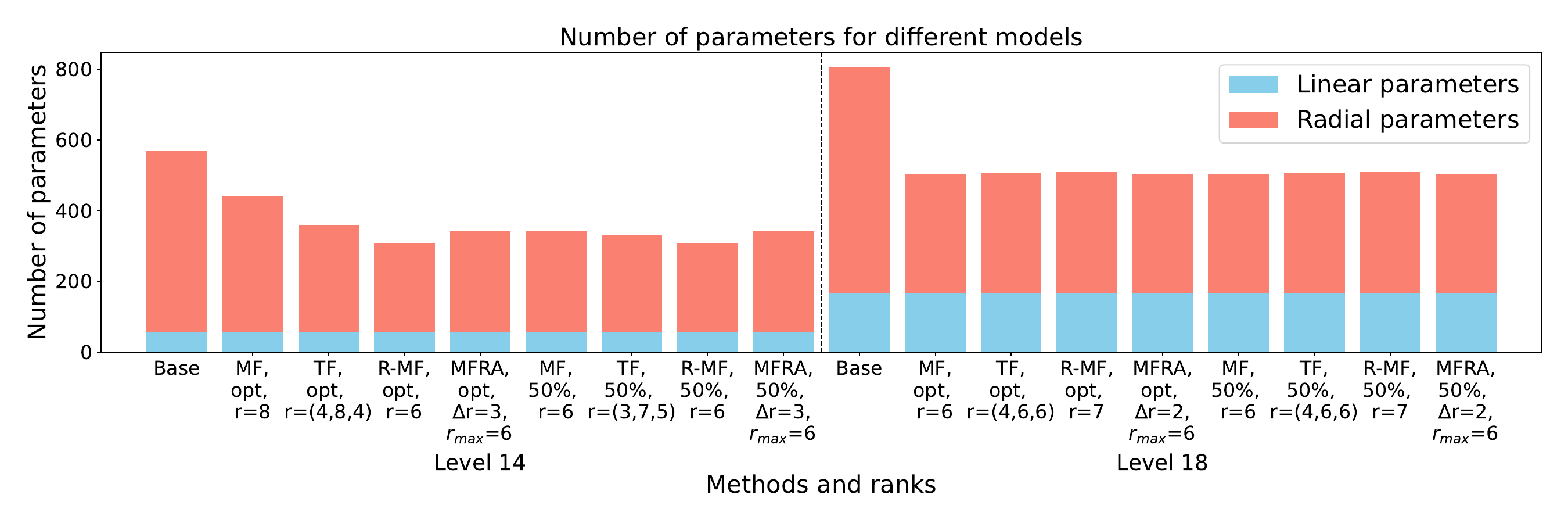}
    \caption{Histogram with the ranks and the number of parameters (linear and radial) for potentials of the 14th and 18th levels fitted on the Mo-Nb-Ta-W training set.}
    \label{fig:monbtaw_14_18_hist}
\end{figure*}

\begin{figure*}[!ht]
    \centering
    \includegraphics[width=\textwidth]{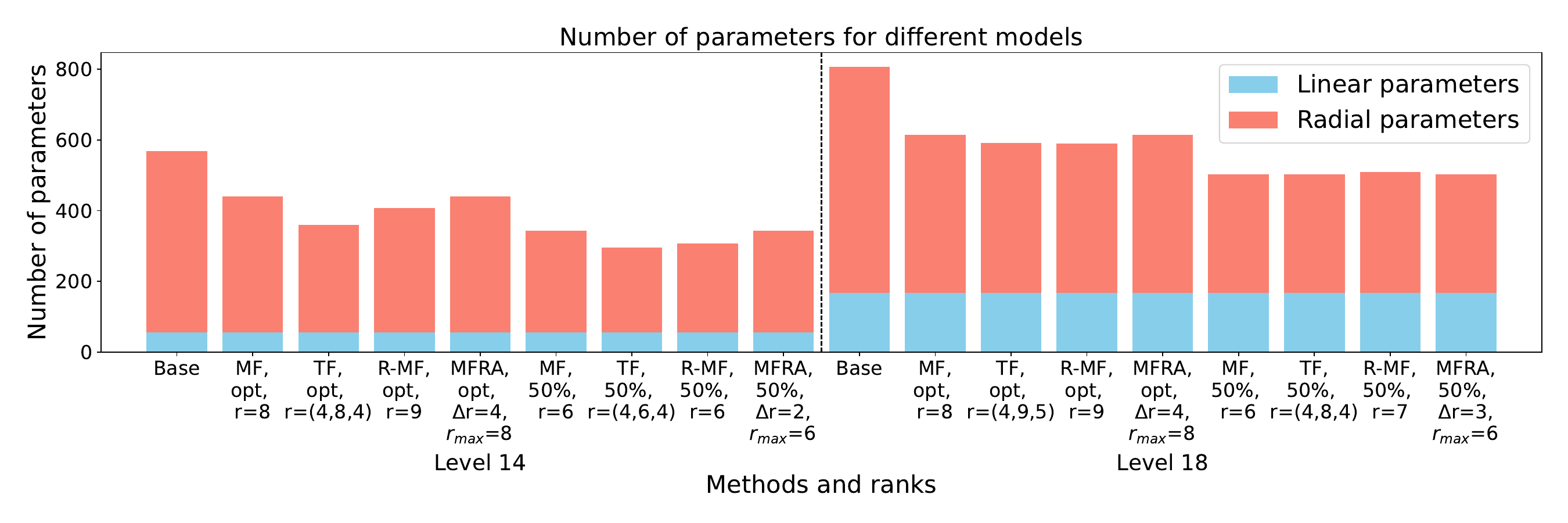}
    \caption{Histogram with the ranks and the number of parameters (linear and radial) for potentials of the 14th and 18th levels fitted on the FLiNaK training set.}
    \label{fig:flinak_14_18_hist}
\end{figure*}

\subsection{Determining the optimal ranks on the example of Mo-Nb-Ta-W}

Here, we discuss the choice of the optimal rank for the MF, R-MF, and TF MTPs. We refer to optimal rank as the rank that yields a reduction (or compression) of not less than 20\% of the radial parameters of the base (non-compressed) MTP and at which the loss function takes its minimum value. We note that the reduction of 20\% of parameters was chosen as a reasonable threshold under which the compression has almost no sense, as the compression of less than 20\% of parameters makes this potential similar to the base potential. We demonstrate the results for the ensembles of MTPs of the 14th level trained on Mo-Nb-Ta-W. 

We start with the MF and R-MF MTPs. The dependence of the loss function calculated on the Mo-Nb-Ta-W validation set on the rank of the factorized matrix of MTP radial parameters is shown in Figure ~\ref{fig:optimal_rank}~(a) (MF) and Figure ~\ref{fig:optimal_rank}~(b) (R-MF). For MF, the loss function significantly decreases up to the rank of 4 and then continues to decrease more gradually to the rank of 8 (which corresponds to 384 radial parameters) and has the minimum value for this rank. The loss function obtained with R-MF demonstrates similar behavior and reaches a minimum for the rank of 6 corresponding to 252 radial parameters. Thus, the optimal ranks for the MF and R-MF MTPs of the 14th level fitted on the Mo-Nb-Ta-W training set are $r=8$ and $r=6$, respectively.

Figure~\ref{fig:optimal_rank}~(c) demonstrates the dependence of the loss function obtained using the TF MTP with the $(r_1,r_2,r_3)$ TT-ranks. The number of radial parameters is given in parentheses. The loss function decreases significantly up to $(r_1,r_2,r_3)=(4,4,4)$ and then flattens. The optimal TT-ranks are $(r_1,r_2,r_3)=(4,8,4)$ which corresponds to 304 radial parameters.

We chose the optimal ranks for the MTPs of other levels fitted on Mo-Nb-Ta-W, as well as for the potentials trained on FLiNaK and glycine in a similar manner. From Figures~\ref{fig:optimal_rank}~(a, b, c) it can be seen that the loss functions obtained for optimal ranks are close to those obtained for the ranks related to compression of 50\%. Therefore, in this work, we also investigate the compression of approximately 50\%. Histograms with ranks and the number of parameters for potentials of the 14th and 18th levels trained on Mo-Nb-Ta-W and FLiNaK are given in Figures ~\ref{fig:monbtaw_14_18_hist} and ~\ref{fig:flinak_14_18_hist}, respectively. We provide similar histograms for the MTPs of the 12th and 16th levels in the supplementary material. We note that we trained MTPs only of the 20th level for glycine and provide the ranks and the number of radial parameters for each MTP in Table ~\ref{tab:glycine_loss_errors} together with the fitting errors.

\subsection{Comparison of the compressed MTPs and ETN for Mo-Nb-Ta-W}

\begin{figure*}[!ht]
    \centering
    \includegraphics[width=0.75\textwidth]{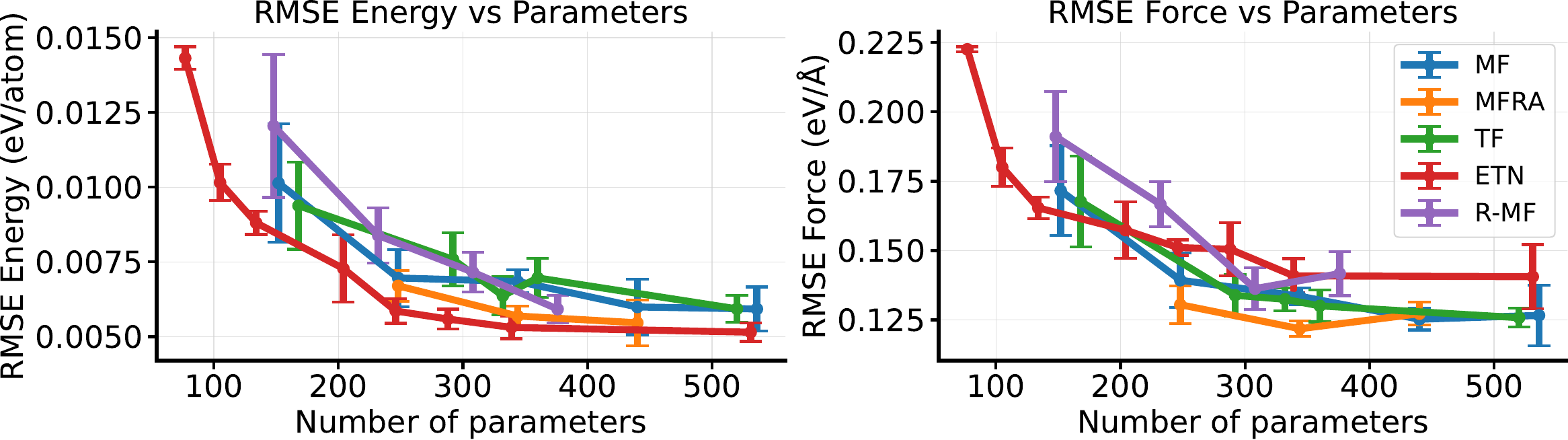}
    \caption{Root mean square errors (RMSEs) for energies and forces calculated with ETN and the compressed MTPs on the Mo-Nb-Ta-W validation set. We provide the results with 68\% confidence interval (i.e., 1-$\sigma$ interval).}
    \label{fig:etn_vs_compressed_mtp}
\end{figure*}

In Figure \ref{fig:etn_vs_compressed_mtp} the dependence of the energy and force validation errors on the number of parameters of the compressed MTPs of the 14th level and ETN fitted on the Mo-Nb-Ta-W training set is demonstrated. We found that the compressed MTPs require a similar number of parameters as ETN to reach similar energy validation error within the statistical error, but most of them give smaller force validation error than ETN. As opposed to Figure 7 in Ref. \cite{hodapp2024-ETN} in which the base MTP was compared to ETN, the compressed MTPs in Figure \ref{fig:etn_vs_compressed_mtp} do not require more parameters to reach similar accuracy as ETN. Thus, in this work we resolved the problem of growing the number of MTP radial parameters. Our focus in further discussion of the results of this work is on the compressed MTPs and we do not make further comparisons with ETN.

\subsection{Loss functions and training errors}

In this subsection, we analyze the values of loss functions and fitting errors calculated with ensembles of different MTPs with the optimal ranks and with the compression of approximately 50\% (i.e., two times reduction) of the number of the radial parameters in the base MTP. In addition, on the example of the FLiNaK system we demonstrate that the tensor of the ACE radial parameters can be compressed with matrix factorization.

\begin{figure*}[!ht]
    \centering
    \includegraphics[width=\textwidth]{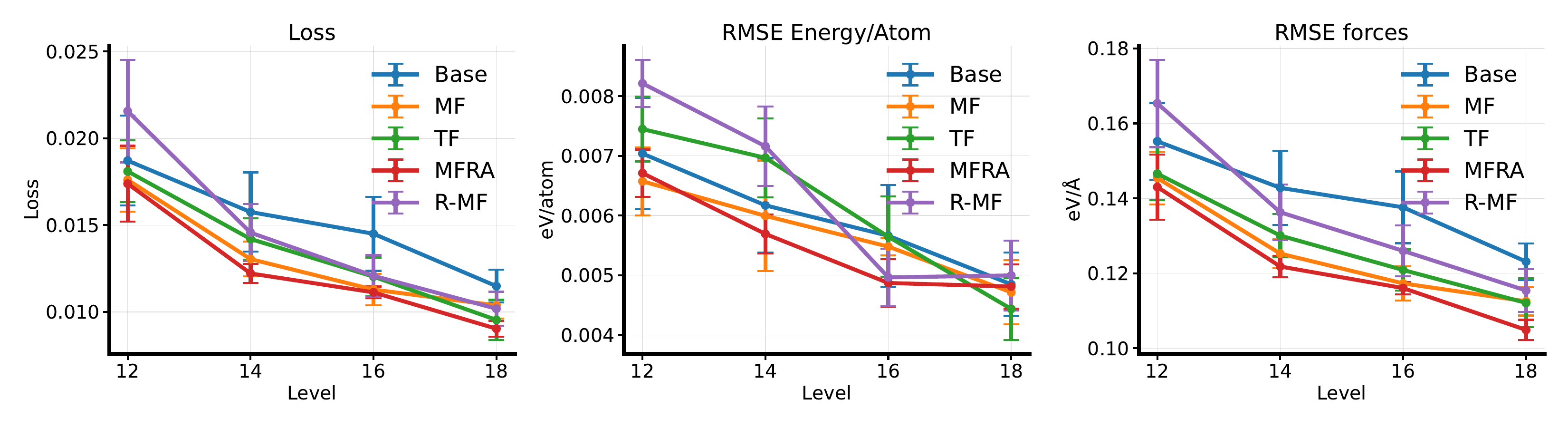}
    \caption{Loss function and root mean square errors (RMSEs) for energies, forces, stresses for different MTPs trained on Mo-Nb-Ta-W with the optimal ranks. The calculations are conducted on the validation set. We provide the results with 68\% confidence interval (i.e., 1-$\sigma$
 interval).}
    \label{fig:monbtaw_errors_best}
\end{figure*}

\begin{figure*}[!ht]
    \centering
    \includegraphics[width=\textwidth]{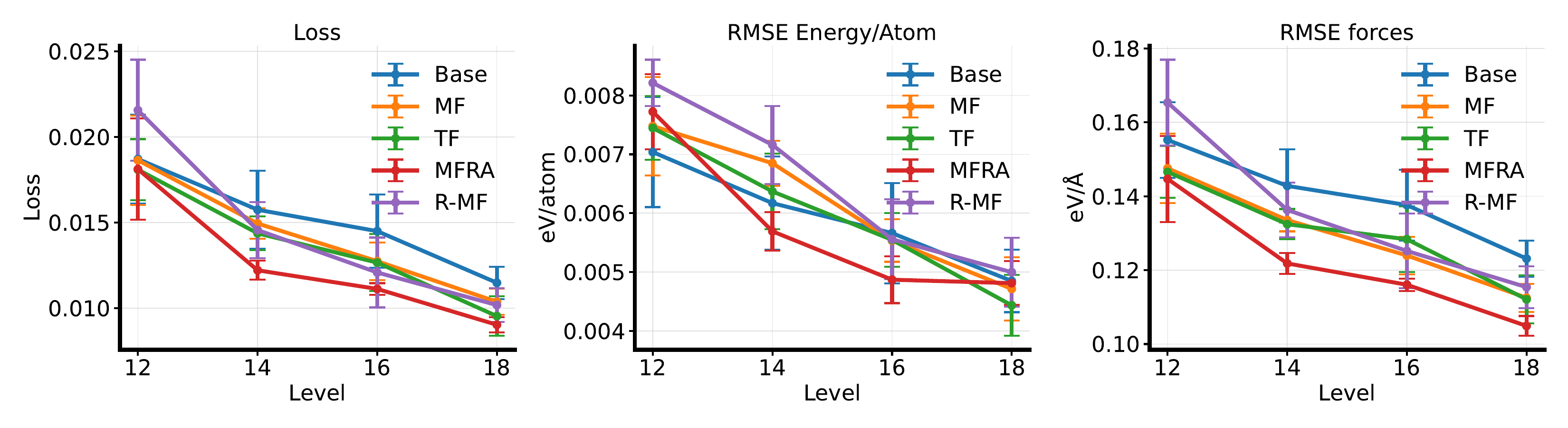}
    \caption{Loss function and root mean square errors (RMSEs) for energies, forces, stresses for different potentials trained on Mo-Nb-Ta-W with 50\% compression. The calculations are conducted on the validation set. We provide the results with 68\% confidence interval (i.e., 1-$\sigma$
 interval).}
    \label{fig:monbtaw_errors_50}
\end{figure*}

The results of fitting of potentials on Mo-Nb-Ta-W are shown in Figures ~\ref{fig:monbtaw_errors_best} and ~\ref{fig:monbtaw_errors_50}. In Figure ~\ref{fig:monbtaw_errors_best} we observe the decrease in loss functions, energy and force fitting errors with the increase of the level of all MTPs. We also see that all the compressed MTPs (MF, MFRA, R-MF, and TF) of the 16th and 18th levels give higher accuracy, i.e., smaller loss functions, energy and force fitting errors than the base MTPs. However, for the MTPs of the 12th level, we conclude that the compressed MTPs give either worse (R-MF) or slightly lower (MF, TF, and MFRA) results compared to the base MTP. This effect is related to the number of linear parameters and the complexity of MTP. There are 29 linear parameters and three radial functions in the MTP of the 12th level, and thus, all the radial parameters presented in the base MTP can be critical for accurate fitting of the compressed MTPs. Furthermore, the MTPs of the 6th and 10th level have two and three radial functions, respectively, and they only have 5 and 16 linear parameters, respectively. As demonstrated in the supplementary material, compression of these potentials gave slightly greater values of the loss functions for the MF MTP at any rank and, thus, there is no sense in compression of the MTPs of these levels from the point of view of accuracy. High-level MTPs of the 16th and 18th levels have 92 and 163 linear parameters, respectively, and four radial functions for each. Therefore, the number of the radial parameters can be excessive and some of them are not necessary for an accurate fitting. In Figure ~\ref{fig:monbtaw_errors_50} only the values for MF with the compression of 50\% became greater than the ones for MF with the optimal ranks. In general, the optimal ranks and the ranks related to the compression of 50\% are close to each other (see Figures ~\ref{fig:optimal_rank}~(a, b, c) and ~\ref{fig:monbtaw_14_18_hist}) for the potentials fitted on Mo-Nb-Ta-W, and therefore the results demonstrated in Figures ~\ref{fig:monbtaw_errors_best} and ~\ref{fig:monbtaw_errors_50} are close to each other. 

It should be noted that the MFRA MTP and the MF MTP require the same CPU time for their fitting. From Table \ref{tab:cpu_time_diff} we see that MF and MFRA require from 540 to 555 CPU minutes to be fitted and the fitting time is particularly independent of rank. The MF MTP and the MFRA MTP require the same CPU time, as we use the same number of BFGS iterations to optimize each of them, and most of the BFGS steps for the MFRA MTP are conducted at $r=r_{\rm max}$ (see Algorithm \ref{alg:inc-rank} for details), which almost corresponds to fitting of the MF MTP at the same, fixed rank. Thus, we do not sacrifice CPU time for accurate training of the MFRA MTP.
\begin{table}[!ht]
\caption{Total CPU time (in minutes) required for fitting of the MF MTP and the MFRA MTP of the 14th level with various ranks on the Mo-Nb-Ta-W system. The results are given for the ensembles of five MF and MFRA with 68\% confidence interval (i.e., 1-$\sigma$ interval). }
\label{tab:cpu_time_diff}
\begin{center}
\begin{tabular}{c|c|c|c|c}
\hline
\hline
 & $r=4$ & $r=6$ & $r=8$ & $r=10$ \\
\hline
MF & $542 \pm 6$ & $543 \pm 9$ & $544 \pm 6$ & $545 \pm 16$ \\
MFRA ($\Delta r = 2$) & $545 \pm 5$ & $552 \pm 2$ & $552 \pm 12$ & $555 \pm 18$ \\
\hline
\hline
\end{tabular}
\end{center}
\end{table}

\begin{figure*}[!ht]
    \centering
    \includegraphics[width=\linewidth]{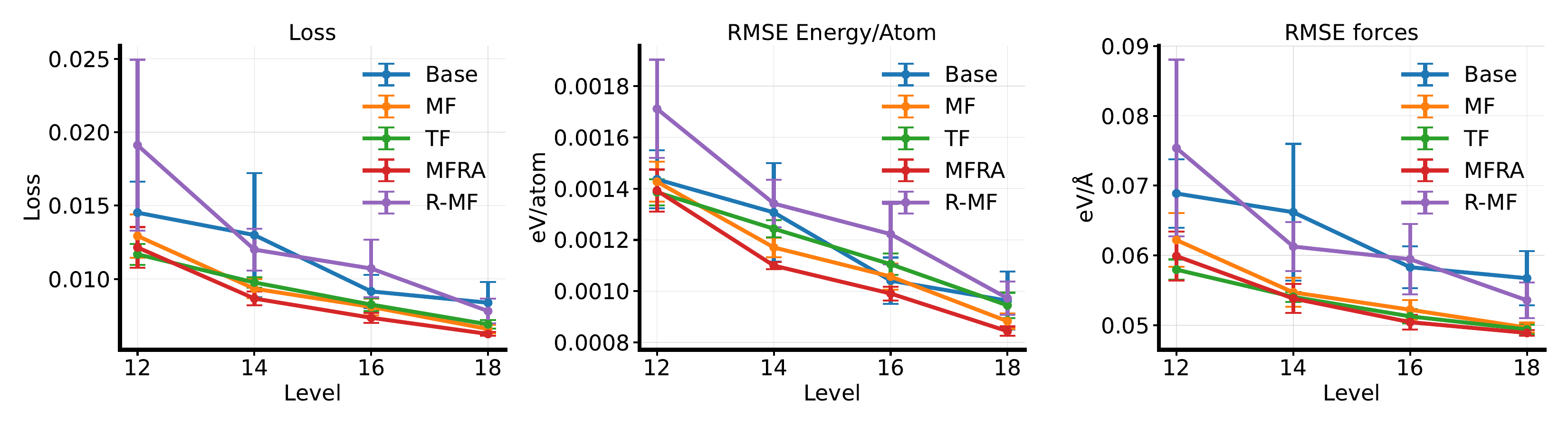}
    \caption{Loss function and root mean square errors (RMSEs) for energies, forces, stresses for potentials trained on FLiNaK with optimal ranks. We provide the results with 68\% confidence interval (i.e., 1-$\sigma$
 interval).}
    \label{fig:flinak_errors_best}
\end{figure*}

\begin{figure*}[!ht]
    \centering
    \includegraphics[width=\textwidth]{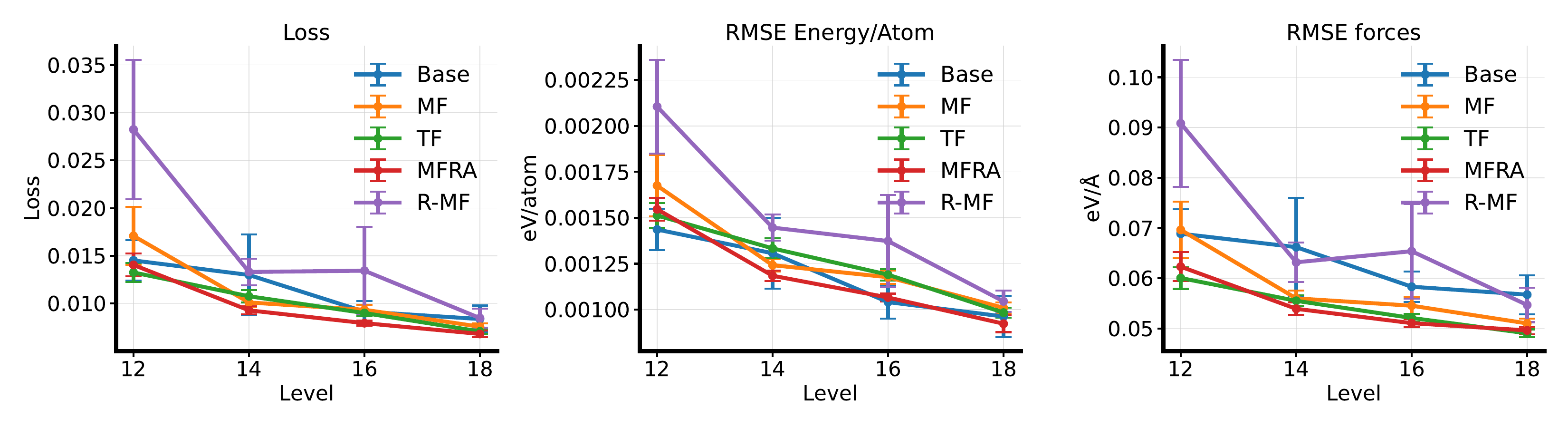}
    \caption{Loss function and root mean square errors (RMSEs) for energies, forces, stresses for potentials trained on FLiNaK with 50\% compression. We provide the results with 68\% confidence interval (i.e., 1-$\sigma$
 interval).}
    \label{fig:flinak_errors_50}
\end{figure*}

In Figures ~\ref{fig:flinak_errors_best} and ~\ref{fig:flinak_errors_50} we see the values of the loss functions and the fitting errors for FLiNaK. From Figure ~\ref{fig:flinak_errors_best} we conclude that the MF, MFRA, and TF MTPs of all levels with the optimal ranks give smaller loss functions and force errors than the base MTPs. However, R-MF demonstrates values similar to the base MTPs and gives greater force fitting error and loss function for the 12th level. Energy fitting errors are statistically similar for all models except for R-MF of the 12th and 16th levels, yielding the worst energy errors. In Figure ~\ref{fig:flinak_errors_50} we observe similar loss functions and energy errors for all MTPs excluding R-MF of the 12th and 16th levels, again demonstrating the worst accuracy among all the models considered. At the same time, force errors are still lower for MF, MFRA, and TF of levels greater than 12 compared to the base MTP even for a two times reduced number of radial parameters.

\begin{figure*}[!ht]
    \centering
    \includegraphics[width=0.66\textwidth]{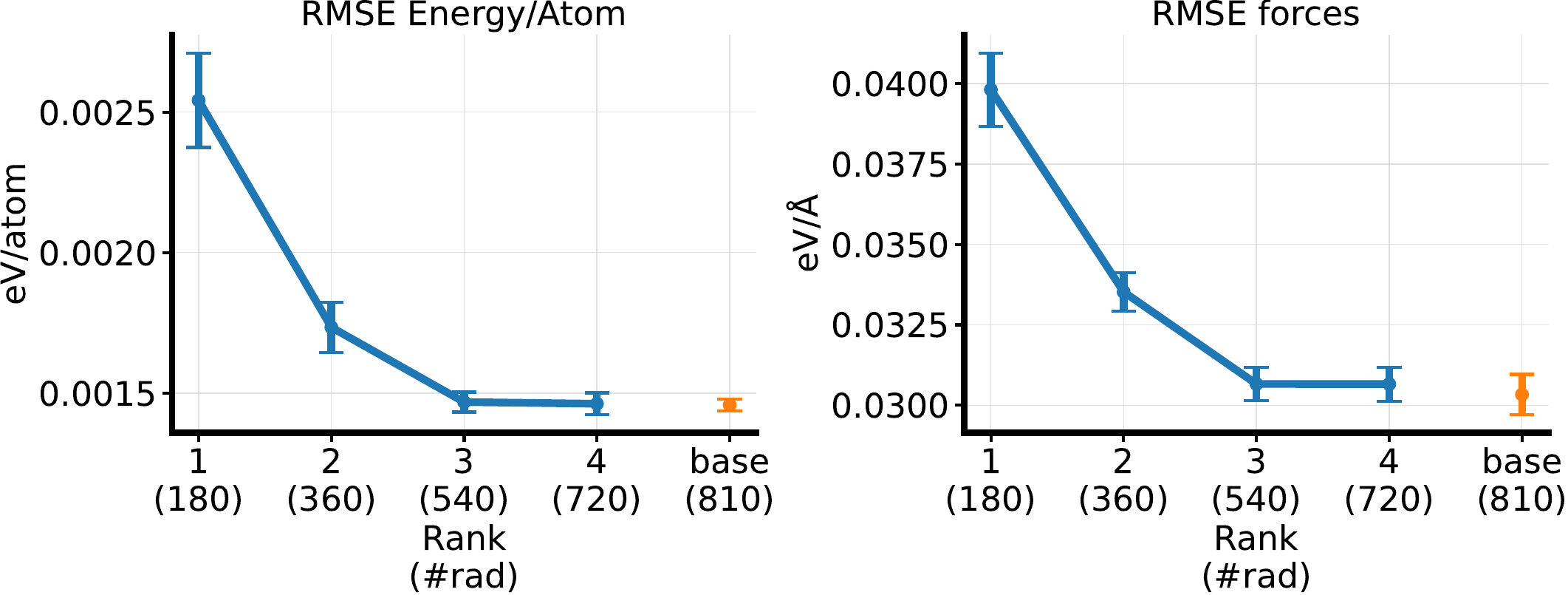}
    \caption{Root mean square errors (RMSEs) for energies and forces for the base and compressed ACE potentials trained on FLiNaK. We provide the results with 68\% confidence interval (i.e., 1-$\sigma$ interval).}
    \label{fig:flinak_ace}
\end{figure*}

In Figure \ref{fig:flinak_ace} the energy and force fitting errors are shown for the ensembles of the base and the compressed ACE fitted on the FLiNaK training set. It could be seen that we do not lose in accuracy when we compress the radial part of ACE and reduce the number of the radial parameters 1.5 times using the factorization of the reshaped matrix of the ACE radial parameters. On the example of the FLiNaK system we see that the proposed methodology for compression of MLIPs is universal and can be applied to various potentials.

Finally, we fitted ensembles of different types of 20th-level MTPs on the training set including glycine molecular crystals. We used the same level of MTP as in Ref. \cite{rybin2025-mol-cryst} and we also took the training set from this paper. The results are shown in Table~\ref{tab:glycine_loss_errors}. We conclude that the potentials with the optimal ranks give slightly smaller loss functions and fitting errors than the potentials with the compression of 50\% of the radial parameters, but the difference is negligible: approximately 0.5 meV/atom for energy, less than 10 meV/\AA ~for forces, and less than 0.1 eV for stresses. We also note that the ensemble of the base MTPs and the ensembles of the compressed potentials with the optimal ranks yield similar loss functions, energy and force errors, and these compressed potentials give slightly smaller stress errors. Thus, as for Mo-Nb-Ta-W and FLiNaK, reducing the number of the radial parameters does not worsen the accuracy of potentials fitted on glycine.

\begin{table}[!ht]
\caption{Loss functions and root mean square errors for energies, forces, and stresses predicted by different MTPs of 20th level (each of them has 288 linear parameters) fitted on glycine. The results are given with 68\% confidence interval (i.e., 1-$\sigma$ interval). Ranks of these potentials and a number of radial parameters for each MTP are also provided.}
\label{tab:glycine_loss_errors}
\begin{center}
\begin{tabular}{c|c|c|c|c|c}
\hline
\hline
MTP & rank & loss & energy error & force error & stress error     \\
& (\# rad.) & 1e-4 & meV/atom & meV/\AA & eV \\
\hline
base & - & 148 $\pm$ 14 & 5.2 $\pm$ 0.4 & 109 $\pm$ 5 & 1.05 $\pm$ 0.08  \\
& (640) &  &  &  &  \\
MF opt & 9 & 134 $\pm$ 6 & 5.1 $\pm$ 0.2 & 104 $\pm$ 2 & 0.95 $\pm$ 0.04  \\
& (504) &  &  &  &  \\
MF 50\% & 6 & 159 $\pm$ 13 & 5.7 $\pm$ 0.4 & 112 $\pm$ 4 & 1.06 $\pm$ 0.07  \\
& (334) &  &  &  &  \\
R-MF opt & 11 & 149 $\pm$ 8 & 5.2 $\pm$ 0.2 & 109 $\pm$ 3 & 1.02 $\pm$ 0.03   \\
& (495) &  &  &  &  \\
R-MF 50\% & 7 & 172 $\pm$ 16 & 5.8 $\pm$ 0.5 & 117 $\pm$ 5 & 1.07 $\pm$ 0.06   \\
& (343) &  &  &  &  \\
MFRA opt & 9 & 130 $\pm$ 4 & 5.0 $\pm$ 0.2 & 102 $\pm$ 2 & 0.98 $\pm$ 0.04  \\
& (504) &  &  &  &  \\
MFRA 50\% & 6 & 146 $\pm$ 8 & 5.4 $\pm$ 0.2 & 108 $\pm$ 3 & 1.01 $\pm$ 0.04  \\
& (334) &  &  &  &  \\
TF opt & 4,10,5 & 129 $\pm$ 6 & 4.9 $\pm$ 0.2 & 101 $\pm$ 2 & 0.92 $\pm$ 0.05  \\
& (466) &  &  &  &  \\
TF 50\% & 4,8,4 & 143 $\pm$ 7 & 5.4 $\pm$ 0.2 & 107 $\pm$ 3 & 0.99 $\pm$ 0.03  \\
& (336) &  &  &  &  \\

\hline
\hline
\end{tabular}
\end{center}
\end{table}

From the figures and the table provided in this subsection, we make several important conclusions. First, compressed MTPs of the 14th level and higher with optimal ranks improve the accuracy of the base MTPs and potentials with the compression of 50\% of the radial parameters do not significantly degrade the accuracy of the base potentials. Second, the accuracy of MTPs with the twice reduced number of the radial parameters is worse, but close to the MTPs with the optimal ranks. Therefore, an MTP with compression of 50\% is a reasonable compromise between the accuracy and the number of parameters. Finally, the MFRA potentials recommended themselves as the most accurate MTPs in terms of the loss function values and fitting errors. Thus, the rank augmentation provided in Algorithm ~\ref{alg:inc-rank} is a useful procedure in terms of the accuracy of fitting.

\begin{table}[!ht]
\caption{Densities (in g/cm$^3$ units) for FLiNaK at different temperatures, calculated with different MTPs. The results are given with 68\% confidence interval, i.e., 1-$\sigma$ interval. The experimental density is taken from Ref.~\cite{romatoski2017fluoride}.} \label{tab:density}
\begin{center}
\begin{tabular}{c|c|c|c|c}
\hline
\hline
Method & \# Radial & 800 K & 1000 K & 1200 K \\ 
& params. &  &  &  \\
\hline
Base & 512 & 2.018 $\pm$ 0.002 & 1.904 $\pm$ 0.005 & 1.803 $\pm$ 0.003 \\
MF opt & 384 & 2.016 $\pm$ 0.003 & 1.905 $\pm$ 0.002 & 1.802 $\pm$ 0.002 \\
MF 50\% & 240 & 2.015 $\pm$ 0.006 & 1.907 $\pm$ 0.006 & 1.807 $\pm$ 0.003 \\
TF opt & 380 & 2.016 $\pm$ 0.003 & 1.906 $\pm$ 0.002 & 1.801 $\pm$ 0.002 \\
TF 50\% & 276 & 2.015 $\pm$ 0.002 & 1.904 $\pm$ 0.002 & 1.798 $\pm$ 0.004 \\
MFRA opt & 384 & 2.013 $\pm$ 0.002 & 1.904 $\pm$ 0.002 & 1.803 $\pm$ 0.003 \\
MFRA 50 \% & 192 & 2.015 $\pm$ 0.004 & 1.905 $\pm$ 0.004 & 1.798 $\pm$ 0.004 \\
R-MF opt & 361& 2.015 $\pm$ 0.003 & 1.906 $\pm$ 0.003 & 1.800 $\pm$ 0.008 \\
R-MF 50 \% & 287 & 2.018 $\pm$ 0.002 & 1.908 $\pm$ 0.004 & 1.798 $\pm$ 0.002 \\
Experiment & - & 2.080 & 1.955 & 1.830 \\
\hline
\hline
\end{tabular}
\end{center}
\end{table}

\begin{figure*}[!ht]
	\centering
	\begin{minipage}[h]{0.5\linewidth}
		\center{\includegraphics[trim={0cm 0cm 0cm 0cm},clip, width=1\linewidth]{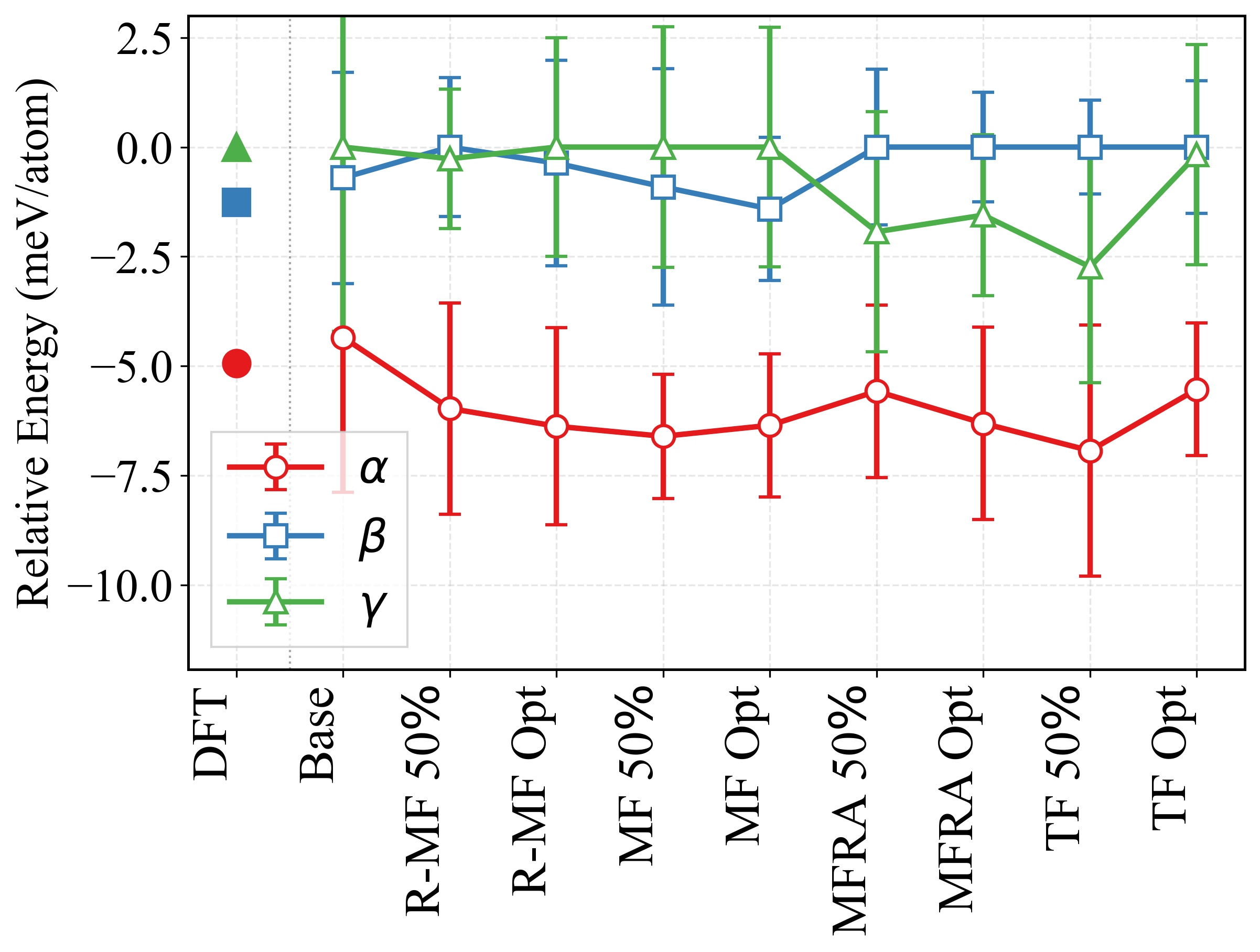}}
	\end{minipage}
    \caption[]{Relative stability of glycine polymorphs calculated using MTP of the 20th level with different compression algorithms and DFT. Data obtained in the scope of DFT is taken from our previous calculations~\cite{rybin2025-mol-cryst}. Averaging is done over 10 separately trained potentials for each model.}
    \label{fig:glycine_stability}
\end{figure*}

\subsection{Temperature dependence of the density of molten FLiNaK}
To examine the predictive power and stability of the compressed MTPs at finite temperature here, we calculated the density of molten LiF-NaF-KF (in an eutectic composition 46.5-11.5-42 mol\% usually labeled as FLiNaK). FLiNaK at temperatures of 800-1200 K. To that end, we conducted molecular dynamics simulations (MD) for 200 picoseconds in the NPT ensemble for the system of 1512 atoms. We utilized the LAMMPS package \cite{thompson2022lammps} for MD simulations. From Tabel \ref{tab:density}, we conclude that all potentials, including those with compression of 50\% of the radial parameters, give similar densities, which are in good agreement with the experimental result~\cite{romatoski2017fluoride} and previous calculations with the base MTP~\cite{rybin2024-FLiNaK}.

\subsection{Energy ranking of glycine polymorphs}

Here, we test whether a compressed version of MTP can can be used for modeling of molecular crystals -- as a test system, we use the glycine molecular crystal, whose polymorphism has been intensively studied (see Refs.~\cite{zhu2012constrained, boldyreva2003structural} and references therein). The three most energetically favorable structures of glycine are polymorphs with space groups $P2_{1}/c$, $P2_{1}$, $P3_{2}$, which are labeled $\alpha$, $\beta$, $\gamma$.

We utilized training datasets obtained in a previous study\cite{rybin2025-mol-cryst}. Our benchmarks with the base and compressed MTPs demonstrate that each MTP (even with the 50\% compression) identified $\alpha$-glycine to have the lowest energy (see Figure~\ref{fig:glycine_stability}), whereas the $\beta$ and $\gamma$ phases have slightly higher energies. This result is consistent with prior DFT-based studies~\cite{zhu2012constrained}, though it contrasts with the experimental relative thermodynamic stability of $\gamma$ > $\alpha$ > $\beta$. This discrepancy is not unexpected and points to the need for more accurate methods of computing intermolecular interaction energies~\cite{chalykh2025momenttensorpotentialequivariant}.

\subsection{Discussion}

The main goal of this work was to verify the methods developed for compression of MLIPs, particularly MTP. We demonstrated the robustness of the proposed methods and showed that even 50\% compression of the radial parameters of MTP retains the accuracy of the base MTP. To calculate the gradients of the loss function with respect to the radial parameters of the compressed MTPs, we used the gradients of the loss function with respect to the radial parameters of the base MTP. Thus, we did not explicitly implement the compressed MTPs, but it can be explicitly implemented in the program code. The profit from the explicit program realization of the compressed form of MTPs is the computational cost of simulations with these potentials. They have fewer parameters than the base MTPs and, therefore, they may require less computational resources for creation of a training set for their fitting. 

\section{Conclusions}

In this study, we proposed methods for reduction of a number of parameters in machine-learning interatomic potentials (MLIPs). To that end, we used two classes of methods: optimization with a fixed-rank constraint including matrix factorization (MF), particularly, skeleton decomposition, tensor factorization (MF) in a form of tensor train decomposition, and a Riemannian version of matrix factorization (R-MF), and optimization with rank augmentation which we applied to MLIP with matrix factorization. 

The methodology is mainly verified using Moment Tensor Potential (MTP) model -- an example of widely used MLIP. We presented all derivations and benchmarked the methodology on three systems: Mo-Nb-Ta-W, molten FLiNaK, and a glycine molecular crystal. The MTP model was selected for testing due to its wide applicability and well-established benchmark performance. Our results showed that even with compression of up to 50\% of the radial parameters of the base (non-compressed) MTP, the fitting errors remain comparable and are sometimes even lower. In principle, as demonstrated, compressing up to 50\% of the radial parameters presents a reasonable compromise between accuracy and the number of parameters, indicating that half of the radial MTP parameters are excessive, at least for high-level potentials.

For the Mo-Nb-Ta-W system, all compressed MTPs of the 16th and 18th levels yielded smaller loss functions and lower force validation errors than the base MTP. However, all compressed MTPs of the 12th level resulted in worse (R-MF) or only marginally lower (MF, TF, and MFRA) loss functions and force validation errors. This is most likely associated with the small number of parameters in the MTP of this level. Energy fitting errors were similar for all MTPs fitted on Mo-Nb-Ta-W. Excluding the MF MTP, all potentials with 50\% compression of the radial parameters yielded results close to the optimal ones for this system.

In the case of FLiNaK, the compressed MTPs are robust, i.e., suitable for large-scale MD simulations, and reproduce the melt densities at finite temperatures with accuracy comparable to the base MTP. In contrast to the Mo-Nb-Ta-W system, compressing 50\% of the radial parameters was not optimal for the MTPs fitted on FLiNaK; consequently, only the force fitting errors for the MF, TF, and MFRA models were smaller than those of the base MTP. In addition to the compressed MTPs, we developed the compressed Atomic Cluster Expansion (ACE) and tested it on the FLiNaK system. It was demonstrated that the compressed ACE requires 1.5 times less radial parameters to be fitted as accurately as the base (original) ACE. Thus, the proposed methodology for reduction of a number of MLIPs parameters is universal and can be applied to different MLIPs.

Finally, the compressed MTPs demonstrate the ability to correctly optimize and rank glycine polymorphs. In this test, we fitted only 20th-level MTPs and found that the base MTP and those with optimal ranks yielded similar or slightly smaller energy RMSEs than the MTPs with 50\% compression. However, the difference is negligible at approximately 0.5~meV/atom.

In the future, we plan to test the active learning algorithm based on D-optimality criterion~\cite{podryabinkin2017_AL} for compressed potentials. This algorithm enable the automated selection of the smallest possible training set without a loss of accuracy or robustness. Our confidence stems from previous work~\cite{novikov2018_RPMD_AL_MTP}, which demonstrated that the size of a training set generated during active learning for MTP depends on its number of parameters, and from the present study, where we show that compressing 50\% of the radial MTP parameters does not significantly affect its accuracy and predictive power.

\section*{Supplementary material}
In the supplementary material, we provide details of Riemannian optimization and additional data.  

\begin{acknowledgments}
This work was supported by the Basic Research Program at the HSE University, Russian Federation. This research was supported in part by computational resources of HPC facilities at the HSE University~\cite{kostenetskiy2021hpc}.

The authors acknowledge Prof. Dr. Alexander Shapeev for providing the code with MTP and python interface and Dmitry Korogod for consultations on the code.
\end{acknowledgments}

\section*{Author declarations}

\subsection*{Conflict of interest}

The authors have no conflicts to disclose.

\subsection*{Author Contributions}

\textbf{Igor Vorotnikov}: Data curation (equal); Formal analysis (equal); Software (equal); Visualization (equal); Writing - original draft (supporting). \textbf{Fedor Romashov}: Data curation (equal); Formal analysis (equal); Software (equal); Visualization (equal); Writing - original draft (supporting). \textbf{Nikita Rybin}: Formal analysis (supporting); Methodology (supporting); Writing - review \& editing (equal). \textbf{Maxim Rakhuba}: Conceptualization (lead); Formal analysis (equal); Methodology (equal); Supervision (equal); Writing - original draft (supporting); Writing - review \& editing (equal). \textbf{Ivan S. Novikov}: Conceptualization (supporting); Formal analysis (equal); Methodology (equal); Software (supporting); Supervision (equal); Writing - original draft (lead); Writing - review \& editing (equal).

\section*{Data Availability Statement}

Data will be made available on request.

%\bibliography{refs}
%

\begin{widetext}
\section*{Supplementary Material: Low-rank matrix and tensor approximations for compression of machine-learning interatomic potentials}

\section*{Details of Riemmanian optimization}

\subsection{Basics of Riemannian manifolds}

For Riemannian manifolds, a tangent space $T_X \mathcal{M}$ is defined at each point $X$, representing a local linearization of the neighborhood around that point, see Figure~1 from the main text for illustration. When the manifold $\mathcal{M}$ is unambiguous, we shorten the notation for the tangent space from $T_X\mathcal{M}$ to simply $T_X$. Formally, the tangent space at $X \in \mathcal{M}_r$ is given by \cite{absil2009optimization}:
\[
\begin{split}
T_X \mathcal{M}_r = & \left\{  U_r A V_r^\top + U_r^\perp B V_r^\top + U_r C (V_r^\perp)^\top \right|
% \\
%  &\ 
 \ \left. A \in \mathbb{R}^{r \times r}, B \in \mathbb{R}^{(m-r) \times r}, C \in \mathbb{R}^{r \times (n-r)} \right\},
\end{split}
\]
where $X = U\Sigma V^\top$ is a full SVD and $U^\perp_r, V^\perp_r$ are defined as in notation section of the main manuscript.

Optimization methods that use the Euclidean gradient can often be generalized using the Riemannian gradient as an analog. In our case, the Riemannian gradient has the following form:
$$\operatorname{grad}~f(X) = \operatorname{Proj}_{T_X} \nabla f(X),$$
which is the projection of the Euclidean gradient onto the tangent space. For any matrix $W \in \mathbb{R}^{m \times n}$, its projection onto $T_{X} \mathcal{M}_r$ is given by:
$$
\operatorname{Proj}_{T_X}(W) = W - P_U^\perp W P_V^\perp,
$$ 
where 
% we use the same notation for the matrix of operator as for the operator itself and
$X = U\Sigma V^\top$, $P_U^\perp = I - U_rU_r^\top$ and $P_V^\perp = I - V_rV_r^\top$.

For optimization, it is sometimes necessary to transfer vectors from one tangent space to another. This requires a vector transport operator. The illustration of the concept is shown on Figure ~\ref{fig:transport}. In this work, we use the operator $\mathcal{T}_{X \to Y}: T_X \mathcal{M}_r \to T_Y \mathcal{M}_r$, chosen such that for all $\xi \in T_X$, $\mathcal{T}_{X \to Y} \xi = \operatorname{Proj}_{T_Y} \xi$.

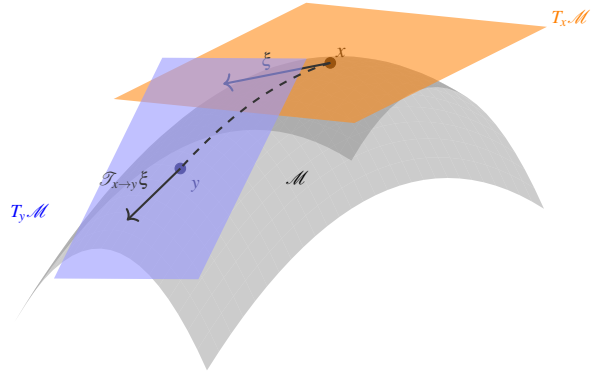
\begin{figure}[!ht]
    \centering
    \input{transport.tikz}
    \caption{Illustration of the concept of vector transport $\mathcal{T}_{x \to y}$ for a smooth manifold~$\mathcal{M}$.}
    \label{fig:transport}
\end{figure}

\subsection{Riemannian BFGS}

To facilitate the representation of linear operators as matrices, it is often convenient to identify $\mathcal{M}_r$ with a manifold in $\mathbb{R}^{mn}$ instead of $\mathbb{R}^{m\times n}$. This is achieved through the vectorization operator $\operatorname{vec}(\cdot)$
\[\operatorname{vec}(A) = \mathrm{reshape}\left({A}, [m_1\cdot m_2\cdot\ldots
\cdot m_k]\right),\]
which provides a natural isomorphism $\mathbb{R}^{m \times n} \cong \mathbb{R}^{mn}$. 
%\mathbb{R}^{n_1 \times n_2 \times \dots \times n_d} \to \mathbb{R}^{n_1 n_2 \dots n_d}$.
Therefore, we sometimes treat points on $\mathcal{M}_r$ as vectors. In this case linear operators $\mathbb{R}^{m \times n} \to \mathbb{R}^{m \times n}$ with some fixed (usually standard) basis can be seen as matrices of shape $mn \times mn$. Under this interpretation, the equality becomes:
$$
\operatorname{vec}\big(\operatorname{Proj}_{T_X}(W)\big) = \big( I_{mn} - P_U^\perp \otimes P_V^\perp \big) \operatorname{vec}(W),
$$
where ``$\otimes$'' is the Kronecker product that is defined as: $$A \otimes B = \begin{bmatrix} a_{11}B & \cdots & a_{1n}B \\ \vdots & \ddots & \vdots \\ a_{m1}B & \cdots & a_{mn}B \end{bmatrix} \in \mathbb{R}^{mp \times nq}.$$
We also used the following well-known identity for the vectorization of a product involving matrices  $A \in \mathbb{R}^{m \times n}$,  $X \in \mathbb{R}^{n \times q}$, and $B \in \mathbb{R}^{p \times q}$:
\begin{equation}\label{eq:vecprod}
    \operatorname{vec}(AXB^\top) = (A \otimes B) \operatorname{vec}(X).
\end{equation}

We apply the BFGS 
algorithm on our Riemannian manifold (hereafter referred to as RBFGS) for the radial parameters $\widehat{C}$ reshaped in matrix form. The algorithm is taken from Ref.~\cite{qi2010riemannian}  with a slight modification -- we add a sufficient descent coefficient $\mu$. In the original paper, $\mu = 0.5$, although it is not shown explicitly. The version we used is presented in Algorithm~\ref{alg:rbfgs}. Here, the optimal values of $\mu$ turn out to be either 0.001 or 0.0001 for different problems.

The RBFGS algorithm, similar to its Euclidean analog, is a quasi-Newton method that builds an approximation of the inverse Hessian from successive gradient evaluations. At each iteration, the Riemannian gradient of the objective is computed and used together with the displacement of iterates to update this approximation. This allows one to incorporate curvature information without explicitly forming the Hessian, which significantly reduces computational cost while still yielding faster convergence in practice.

In more detail, the main steps of the RBFGS algorithm are as follows. First, on $k$-th iteration at the current point $\Theta_k$ on the manifold, we compute the Riemannian gradient of the objective function $L$. This gradient is then multiplied by the current approximation of the inverse Hessian $H_k$ to produce a search direction $\eta_k$. Such a transformation improves the choice of direction compared to plain gradient descent, while still ensuring that the direction remains within the tangent space of the current point. Next, a line search is performed along this direction to find a step of appropriate length $\alpha_k$ that noticeably decreases the objective function at point $R_{\Theta_k}(\alpha_k\eta_k)$. Here, the retraction operator is used to map the tangent vector $\alpha_k \eta_k$ back onto the manifold. After the step is taken and the point is updated on the manifold, the algorithm proceeds with an efficient update rules that adjust the inverse Hessian approximation, so that curvature information from the new iterate is incorporated. For these updates both transport operator $\mathcal{T}_{\Theta_k \to \Theta_{k+1}}$ and inverse transport operator $\mathcal{T}_{\Theta_k \to \Theta_{k+1}}^{-1}$ are used to work with vectors lying in tangent spaces at different points.

The algorithm is universal and works with arbitrary manifolds. However, we need to understand its explicit form for the manifold $\mathcal{M}_r$ of the fixed-rank matrices we work with. All operators used in the algorithm have an efficiently computable explicit form, except for the inverse transport operator. Therefore, the main problem is to derive the inverse transport operator in the matrix form. Next we describe the final formulas and provide their derivation in supplementary material.

\begin{algorithm}[ht]
\caption{RBFGS}
\label{alg:rbfgs}
\KwIn{Riemannian manifold $\mathcal{M}$ with Riemannian metric $g$; 
vector transport $\mathcal{T}$ on $\mathcal{M}$ with associated retraction $R$; 
smooth function $L({\bm \Theta})$ on $\mathcal{M}$; initial iterate ${\bm \Theta}_0 \in \mathcal{M}$; 
initial approximation of inverse Hessian $H_0$; coefficient of sufficient decrease $\mu$.}

\For{$k = 0, 1, 2, \dots$}{
    Obtain $\eta_k \in T_{{\bm \Theta}_k} \mathcal{M}$ by solving 
    $\eta_k = -H_k \,\mathrm{grad}\, L({\bm \Theta}_k)$\;
    
    Set step size $\alpha \gets 1$, $c \gets g(\mathrm{grad}\, L({\bm \Theta}_k), \eta_k)$\;
    
    \While{$L(R_{{\bm \Theta}_k}(2\alpha \eta_k)) - L({\bm \Theta}_k) < 2 \mu \alpha c$}{
        $\alpha \gets 2\alpha$\;
    }
    
    \While{$L(R_{{\bm \Theta}_k}(\alpha \eta_k)) - L({\bm \Theta}_k) \geq \mu \alpha c$}{
        $\alpha \gets 0.5\alpha$\;
    }
    
    Set ${\bm \Theta}_{k+1} \gets R_{{\bm \Theta}_k}(\alpha \eta_k)$\;
    
    Define $s_k \gets \mathcal{T}_{\Theta_k \to \Theta_{k+1}}(\alpha \eta_k)$\;
    
    Define $y_k \gets \mathrm{grad}\, L({\bm \Theta}_{k+1}) 
    - \mathcal{T}_{\Theta_k \to \Theta_{k+1}} (\mathrm{grad}\, L({\bm \Theta}_k))$\;
    
    Define operator $H_{k+1}: T_{{\bm \Theta}_{k+1}} \mathcal{M} \to T_{{\bm \Theta}_{k+1}} \mathcal{M}$ by
    
    \Indp
    \begin{align*}
    &H_{k+1} p = \widetilde{H}_k p 
      - \frac{g(y_k, \widetilde{H}_k p)}{g(y_k, s_k)} s_k
      - \frac{g(s_k, p)}{g(y_k, s_k)} \widetilde{H}_k y_k +\\
      &\quad + \frac{g(s_k, p)\, g(y_k, \widetilde{H}_k y_k)}{g(y_k, s_k)^2} s_k
      + \frac{g(s_k, s_k)}{g(y_k, s_k)} p, 
      \quad \forall p \in T_{{\bm \Theta}_{k+1}} \mathcal{M};
    \end{align*}
    \Indm
    
    where 
    $\widetilde{H}_k = \mathcal{T}_{\Theta_k \to \Theta_{k+1}} \circ H_k \circ (\mathcal{T}_{\Theta_k \to \Theta_{k+1}})^{-1}$\;
}
\end{algorithm}

% We start from reshaping the tensor of the radial parameters $C$: 
% $$
% \widehat{C} \;=\;\mathrm{reshape}\bigl(C,\;[N_T\cdot N_T, N_f\cdot N_b]\bigr)
% \;\in\;\mathbb{R}^{(N_T^2)\times (N_fN_b)}.
% $$

% We denote $m = N_T^2$, $n = N_fN_b$, and $\operatorname{rank}(\widehat{C}) = r$. 

We additionally note that the manifold $\mathcal{M}$ in the algorithm in our case is $\mathbb{R}^{|\Xi|} \times \mathcal{M}_r$ (the first term corresponds to the linear parameters, the second term corresponds to the reduced number of the radial parameters) where $r$ is the matrix rank, $\mathcal{M}_{r}$ is a manifold of matrices of this rank. Therefore, we refer to MTP fitted with Algorithm~\ref{alg:rbfgs} as the R-MF MTP. We note that for the ease of presentation we only focus on the optimization on $\mathcal{M}_r$.

Let $z \in \mathcal{M}$ where $\mathcal{M}$ is an arbitrary manifold. By $T_z$ denote the tangent space $T_z \mathcal{M}$ and by 
$P_z$ denote the matrix of orthogonal projector $\operatorname{Proj}_{T_z}$ onto $T_z$ in standard basis.

\begin{Proposition} \label{lem:inv_transport}
Let $\mathcal{M}\subseteq \mathbb{R}^N$ be a Riemannian manifold of the dimensionality $M$.
% Denote $T_x, T_y$ as the tangent spaces at these points.
Let $B_x, B_y \in \mathbb{R}^{N \times M}$ be matrices of orthogonal bases in $T_x$ and $T_y$ respectively, and the operator of vector transport  $\mathcal{T}_{x \to y}$ be a restriction of $\operatorname{Proj}_{T_y}$ on $T_x$.
Then the operator of inverse transport $\mathcal{T}_{x \to y}^{-1}: T_y\mathcal{M} \to T_x\mathcal{M}$ can be expressed in matrix form in standard basis in $\mathbb{R}^N$ as 
$B_x (B_y^\top P_y B_x)^{-1} B_y^\top$.
\end{Proposition}

\begin{proof} \label{proof:inv_tr_common}
Note that $ \mathcal{T}_{x \to y} = \operatorname{Proj}_{T_y} \big|_{T_x}$
and $\mathcal{T}_{x \to y}$ is bijective since the dimensions of tangent spaces at each manifold point are equal and we assume the mapping is surjective. Our goal is to obtain $\mathcal{T}_{x \to y}^{-1}$ in matrix form $\mathbb{R}^{N \times N}$ to work with vectors in the standard basis in $\mathbb{R}^N$.

We propose the following solution. Let matrix $P_{x \to y} \in \mathbb{R}^{M \times M}$ be the mapping that takes a vector from $T_x$ in $B_x$ basis coordinates and outputs in $B_y$ basis coordinates.

Derive the formula for $P_{x \to y}$. On the one hand, for each vector in $T_x $ in $B_x$ coordinates we can multiply it by $P_{x \to y}$ and get its projection on $T_y$ in $B_y$ coordinates. On the other hand, we can multiply it by $B_x$ to transform into standard coordinates in $\mathbb{R}^N$, then project that vector on $T_y$ multiplying by $P_y$ and then change the basis to $B_y$ multiplying by $B_y^+$. Hence, we have the equality:
$$
P_{x \to y} = B_y^+ P_y B_x = B_y^\top P_y B_x,
$$
where the left matrix has size $M \times M$. Therefore:
$$
P_{x \to y}^{-1} = (B_y^\top P_y B_x)^{-1}
$$
However, we need $\mathcal{T}_{x\to y}^{-1}$, not $P_{x \to y}^{-1}$. To obtain this, for each vector in $T_y$ in standard coordinates we have to convert the vector to $B_y$ coordinates, apply $P_{x \to y}^{-1}$, then return the result to standard basis from $B_x$ basis. Hence, the final matrix:
$$
\mathcal{T}_{x \to y}^{-1} = B_x (B_y^\top P_y B_x)^{-1} B_y^\top
$$

For some manifolds, we can further accelerate obtaining the inverse transport matrix by considering explicit formulas for obtaining the basis and orthogonal projector on the tangent space. It can be noted that $B_x$ do not have to be orthogonal basis for formula to work, though it is a good practice to use orthogonal $B_x$ for numerical stability.
\end{proof}

Now consider a specific case in which we are interested in this work: the manifold $\mathcal{M}_r$ of fixed rank $r$ matrices of size $m \times n$. As we mentioned earlier, sometimes we think of matrices as of vectors implying natural isomorphism $\operatorname{vec}(\cdot): \mathbb{R}^{m \times n} \to \mathbb{R}^{mn}$. Let $\tilde{\mathcal{M}}_r$ be a manifold of vectorized matrices from $\mathcal{M}_r$. For matrices $X, Y$ we denote $x = \operatorname{vec}(X),~y = \operatorname{vec}(Y)$.

\begin{Proposition}
  Let $X, Y \in \mathcal{M}_r$ be matrices of shape $m \times n$, with full SVD $X = U_X \Sigma_X V_X^\top$, $Y = U_Y \Sigma_Y V_Y^\top$. The matrix form of the inverse transport operator on $\tilde{\mathcal{M}}_r$ can be expressed as 
  \[
  \begin{split}
  &\mathcal{T}_{x \to y}^{-1} = E^\top \left( (U_Y^\top U_X) \otimes (V_Y^\top V_X) - (U_Y^\top P_{U_Y}^{\perp} U_X) \otimes (V_Y^\top P_{V_Y}^{\perp} V_X) \right) E,
  \end{split}
  \]
  where $P_{U_Y}^{\perp} = I_m - (U_Y)_r (U_Y)_r^\top, ~P_{V_Y}^{\perp} = I_n - (V_Y)_r (V_Y)_r^\top$ and $E$ is a matrix consisting only of 0 and 1 such that Multiplying by this matrix  ``cuts out'' special columns of the initial matrix. 
\end{Proposition}

\begin{proof}  \label{proof:inv_tr_Mr}
For $\mathcal{M}_r$, the explicit form of matrices in $T_X \mathcal{M}_r$\cite{steinlechner2016riemannian} is:  
$$
T_X \mathcal{M}_r = \left\{ \begin{bmatrix} U_{X, r} & U^{\perp}_{X, r} \end{bmatrix}  
\begin{bmatrix} * & * \\ * & 0 \end{bmatrix}  
\begin{bmatrix} V_{X, r} & V^{\perp}_{X, r} \end{bmatrix}^\top \right\},
$$
where $
X = U_X \Sigma_X V_X^\top
$ -- full SVD and $U_{X, r} = (U_X)_r, ~V_{X, r}=(V_X)_r$.
Show that the set of matrices lying in $T_X \mathcal{M}_r$ and having a form $U_X E_{ij} V_X^\top$, where $i \leq r$ or $j \leq r$, is orthogonal:  
$$
\langle U_X E_{i_1 j_1} V_X^\top, U_X E_{i_2 j_2} V_X^\top \rangle_F = \langle E_{i_1 j_1}, E_{i_2 j_2} \rangle_F 
= \begin{cases}
1, & \text{if } i_1 = i_2, j_1 = j_2, \\
0, & \text{otherwise.}
\end{cases}
$$
Notice that the cardinality of this set is $(n+ m - r)r$. It equals dimensionality of $T_X \mathcal{M}_r$, therefore, $\{U_X E_{i j} V_X^\top \mid i \leq r~ \text{or}~ j \leq r\}$ is an orthogonal matrix basis in $T_X \mathcal{M}_r$ and $\{\operatorname{vec}(U_X E_{i j} V_X^\top) \mid i \leq r~ \text{or}~ j \leq r\}$ is an orthogonal vector basis in $T_x \tilde{\mathcal{M}}_r$. To obtain $B_x$ we have to stack these vectors:
$$
B_x =
\begin{bmatrix}
\text{vec}(U_X E_{11} V_X^\top), \text{vec}(U_X E_{12} V_X^\top), \dots
\end{bmatrix}
$$

Since
$$
\text{vec} (U_X E_{ij} V_X^\top) = (U_X \otimes V_X) \text{vec} (E_{ij}),
$$
then  
$$
B_x = (U_X \otimes V_X)
\begin{bmatrix}
\text{vec}(E_{11}), \text{vec}(E_{12}), \dots
\end{bmatrix}
$$

Let us call the matrix in brackets $E$. The explicit matrix form of the projector onto~$T_x \tilde{\mathcal{M}}_r$:
$$
P_x = I_{mn} - (I_m - U_{X, r} U_{X,r}^\top) \otimes (I_n - V_{X, r} V_{X, r}^\top)
$$
Recall that we denote $P_{U_X}^{\perp} = (I_m - U_{X, r} U_{X,r}^\top), ~P_{V_X}^{\perp} = (I_n - V_{X, r} V_{X, r}^\top)$.

Putting it all together and using the Lemma ~\ref{lem:inv_transport}:
\begin{align*}
P_{x \to y} =& B_y^\top P_y B_x = E^\top (U_Y^\top \otimes V_Y^\top)
(I_{mn} - P_{U_Y}^{\perp} \otimes P_{V_Y}^{\perp})
(U_X \otimes V_X) E = \\
= &E^\top \left( (U_Y^\top U_X) \otimes (V_Y^\top V_X) - (U_Y^\top P_{U_Y}^{\perp} U_X) \otimes (V_Y^\top P_{V_Y}^{\perp} V_X) \right) E.
\end{align*}
\end{proof}

\begin{Note}
Notice that since $E$ consists of $M$ columns of the form $e_i$
we can multiply it in $O(\text{\#elements in multiplied matrix})$ time.
\end{Note}

Let
$$
\text{mask} = \text{vec} \left(
\begin{bmatrix}
\mathbf{1}_{r \times r} & \mathbf{1}_{r \times (n -r)} \\
\mathbf{1}_{(m - r) \times r} & \mathbf{0}_{(m - r) \times (n - r)}
\end{bmatrix}
\right) \in \mathbb{R}^N,
$$
where $\mathbf{1}_{n_1 \times n_2}$ is a matrix of size $n_1 \times n_2$ consisting entirely of ones, then for any matrix~$A$
$$
A E = A[:, \text{mask}],
$$
meaning that the new matrix is obtained from $A$ by cutting out the columns which indices $i$ satisfy $\text{mask}[i] = 1$.

\section*{Gradients of the loss function with respect to parameters of tensor train}

Gradients of the loss function $L(\Xi,G^{(1)},G^{(2)},G^{(3)},G^{(4)})$ with respect to all tensor train cores $G^{(k)}, k=1,\ldots,4$:

\[ \frac{\partial L}{\partial G^{(1)}_{1, z_i, \alpha_1}} = 
\sum_{z_j=1}^{N_T}\sum_{\mu=1}^{N_f}\sum_{\beta=1}^{N_b}\sum_{\alpha_2=1}^{r_2}\sum_{\alpha_3=1}^{r_3}\frac{\partial L}{\partial c^{(\beta)}_{z_i, z_j, \mu}}  G^{(2)}_{\alpha_1,z_j,\alpha_2}
G^{(3)}_{\alpha_2,\mu,\alpha_3}
G^{(4)}_{\alpha_3,\beta,1}; \]

\[ \frac{\partial L}{\partial G^{(2)}_{\alpha_1,\,z_j,\,\alpha_2}} = 
\sum_{z_i=1}^{N_T}\sum_{\mu=1}^{N_f}\sum_{\beta=1}^{N_b}\sum_{\alpha_3=1}^{r_3}\frac{\partial L}{\partial c^{(\beta)}_{z_i, z_j, \mu}}  G^{(1)}_{\,1,\,z_i,\,\alpha_1}
G^{(3)}_{\alpha_2,\,\mu,\,\alpha_3}
G^{(4)}_{\alpha_3,\,\beta,\,1}; \]

\[ \frac{\partial L}{\partial G^{(3)}_{\alpha_2,\,\mu,\,\alpha_3}} = 
\sum_{z_i=1}^{N_T}\sum_{z_j=1}^{N_T}\sum_{\beta=1}^{N_b}\sum_{\alpha_1=1}^{r_1}\frac{\partial L}{\partial c^{(\beta)}_{z_i, z_j, \mu}}  G^{(1)}_{\,1,\,z_i,\,\alpha_1}
G^{(2)}_{\alpha_1,\,z_j,\,\alpha_2}
G^{(4)}_{\alpha_3,\,\beta,\,1}; \]

\[ \frac{\partial L}{\partial G^{(4)}_{\alpha_3, \beta, 1}} = 
\sum_{z_i=1}^{N_T}\sum_{z_j=1}^{N_T}\sum_{\mu=1}^{N_f}\sum_{\alpha_1=1}^{r_1}\sum_{\alpha_2=1}^{r_2}\frac{\partial L}{\partial c^{(\beta)}_{z_i, z_j, \mu}} G^{(1)}_{1,z_i,\alpha_1} G^{(2)}_{\alpha_1,z_j,\alpha_2}
G^{(3)}_{\alpha_2,\mu,\alpha_3}. \]

\section*{Histograms with ranks and parameters for potentials of the 12th and 16th levels}

\begin{figure*}[!ht]
    \centering
    \includegraphics[width=\textwidth]{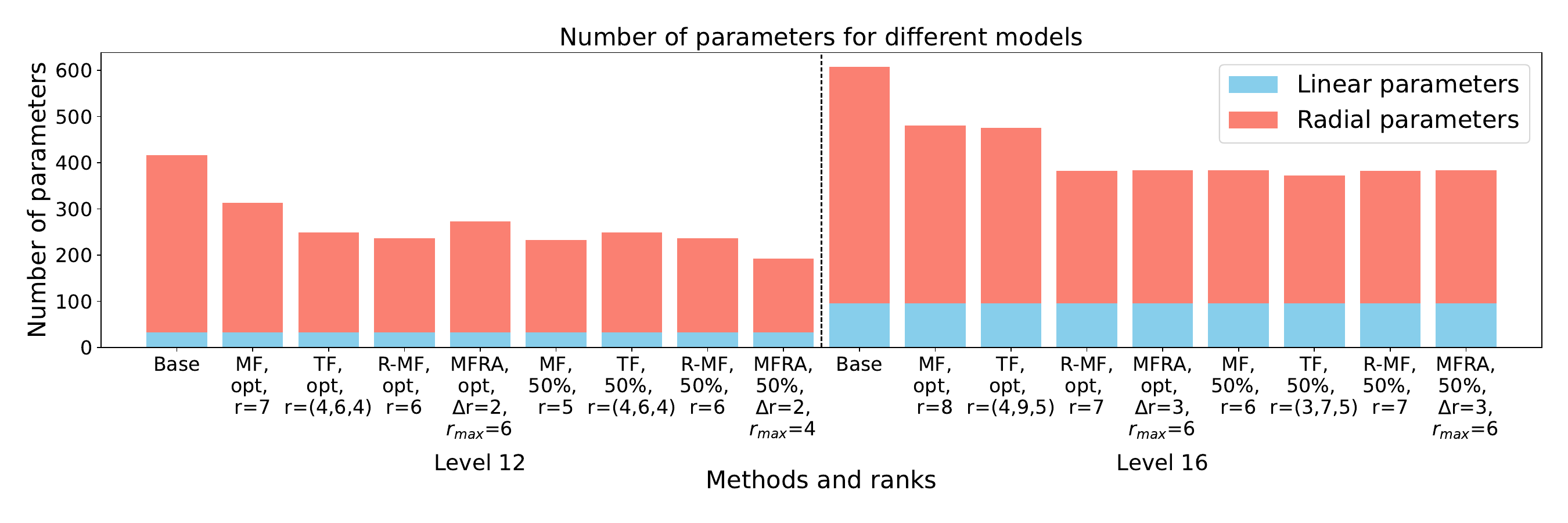}
    \caption{Histogram with the ranks and the number of parameters (linear and radial) for potentials of the 12th and 16th levels fitted on the Mo-Nb-Ta-W training set.}
    \label{fig:monbtaw_12_16_hist}
\end{figure*}

\begin{figure*}[!ht]
    \centering
    \includegraphics[width=\textwidth]{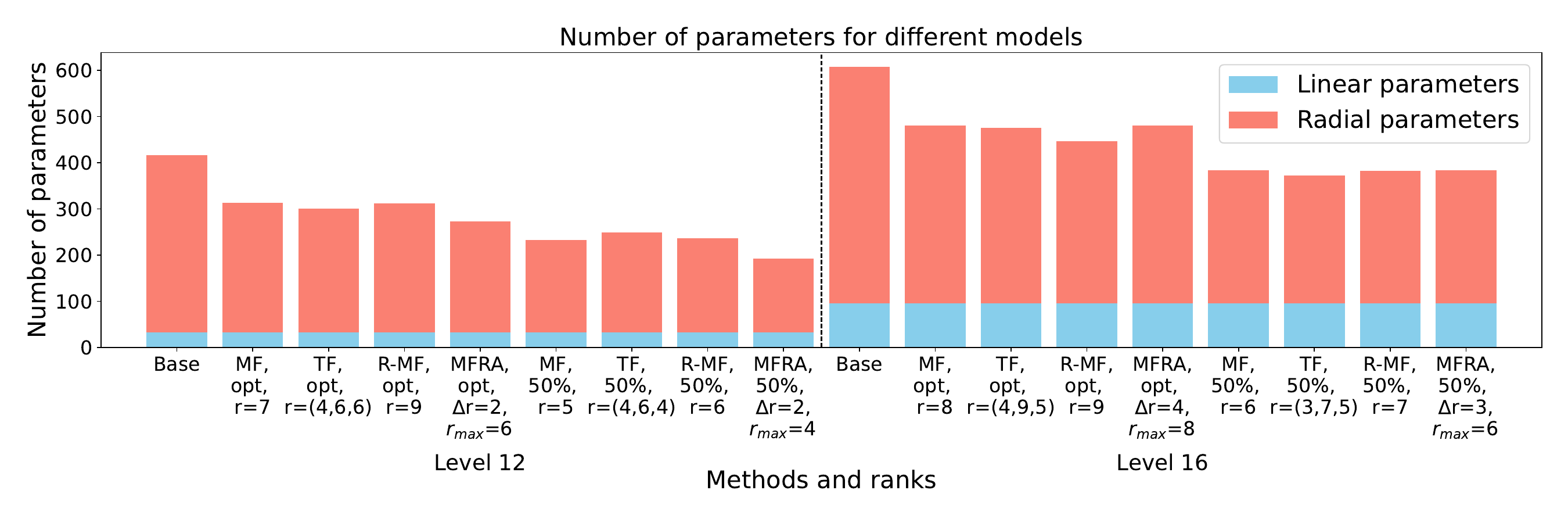}
    \caption{Histogram with the ranks and the number of parameters (linear and radial) for potentials of the 12th and 16th levels fitted on the FLiNaK training set.}
    \label{fig:flinak_12_16_hist}
\end{figure*}

\newpage

\section*{MF MTP of the 6th and 10th level}

From Figure \ref{fig:MF_MTP_6_10_level} we see that the values of the loss function of the MF MTPs are greater than the loss function of the base MTP. Thus, it has no sense to compress MTPs of these levels from the point of view of accuracy.

\begin{figure}[!ht]
    \centering
    \begin{minipage}[h]{0.33\linewidth}
\includegraphics[trim={0cm 0cm 0cm 0cm},clip, width=1\linewidth]{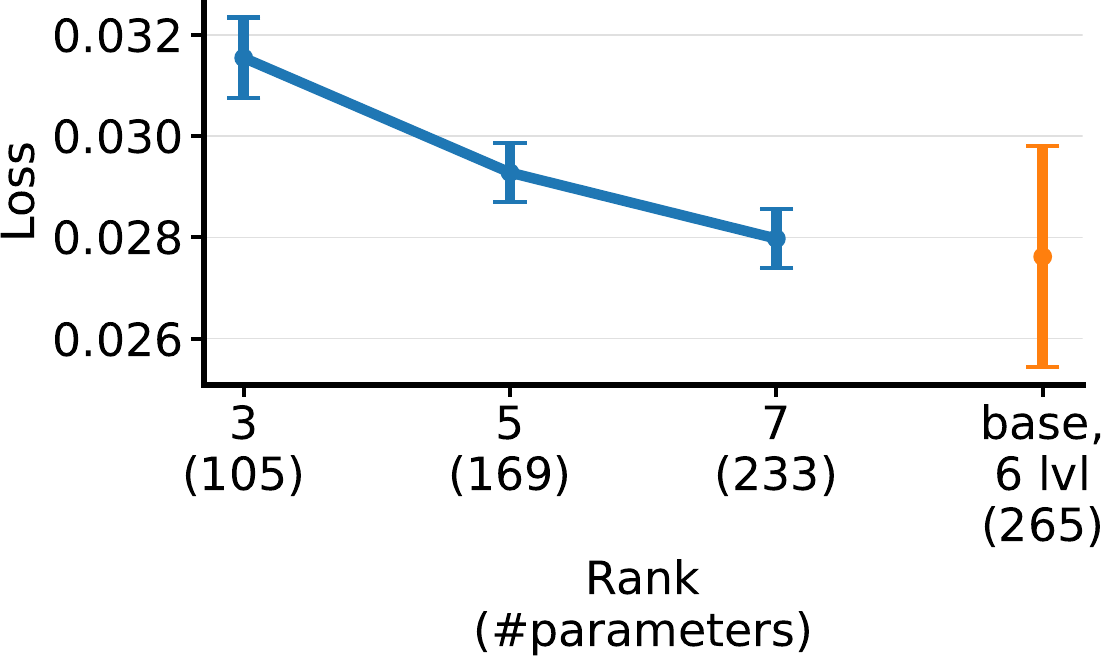}\\
    (a) 
    \end{minipage}
    \begin{minipage}[h]{0.33\linewidth}
\includegraphics[trim={0cm 0cm 0cm 0cm},clip, width=1\linewidth]{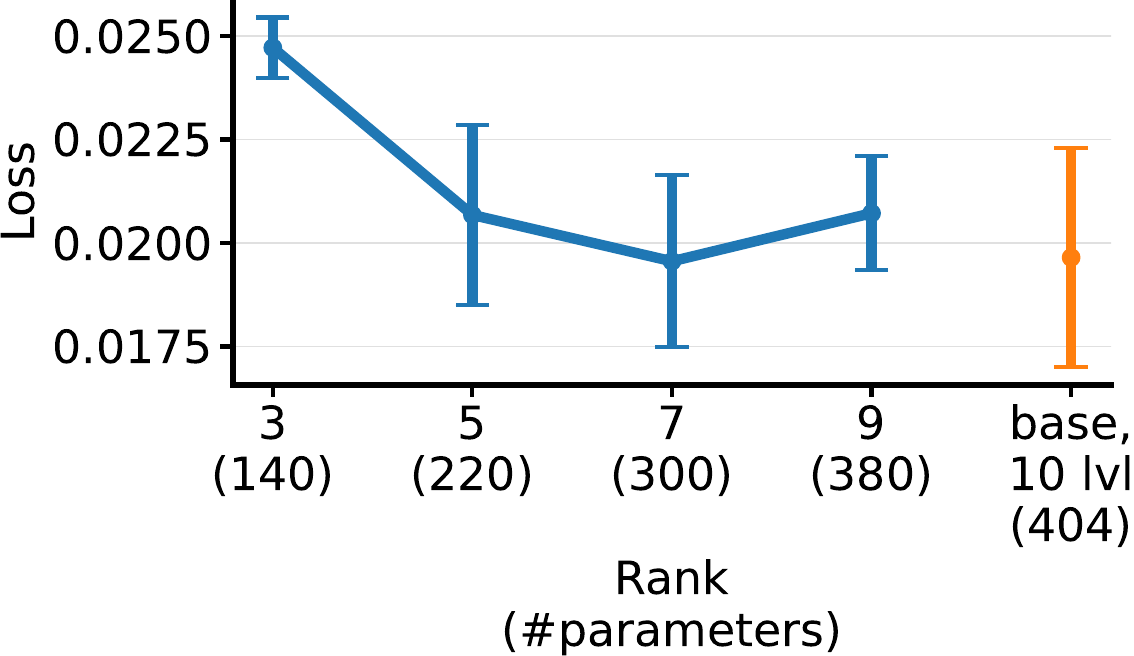}\\
    \small (b) 
    \end{minipage}

    \caption{Dependence of the loss function calculated on the Mo-Nb-Ta-W validation set on the rank of the MF MTP of (a) the 6th level and of (b) the 10th level. Error bars indicate 1-$\sigma$ confidence intervals.}
  \label{fig:MF_MTP_6_10_level}

\end{figure}

\section*{Recommendations for choice of MFRA hyperparameters}

\paragraph*{Setup.} We study MFRA training on the Mo-Nb-Ta-W training set for potentials of the 14th level by varying three hyperparameters: (1) the augmentation interval $s$, (2) the initial rank $r_{\min}$, and (3) the rank increment $\Delta r$. Unless otherwise noted, we use a total budget of 2000 BFGS steps with $\Delta r=2, s=80, r_{\min}=2$ and provide the value of the loss function on the validation set.

\paragraph*{Findings.}
\begin{enumerate}
  \item \textbf{Step interval $s$.} The loss function is broadly insensitive to $s$ across a wide range, with a shallow minimum around $s\in[60,100]$ for the 2000-step budget (Figure~\ref{fig:hyps}a). Too small $s$ (e.g., $\leq 20$) induces MF-like behavior because the ranks grow before the optimizer adapts the low-rank subspace. In contrast, a very large $s$ ineffectively uses the computing budget at higher ranks. 
  \item \textbf{Initial rank $r_{\min}$.} The initial rank $r_{\min}=1$ consistently underperforms higher ranks (Figure~\ref{fig:hyps}b). Choosing $r_{\min}\in\{2,3,4\}$ yields smaller and similar values of the loss functions. An excessively large $r_{\min}$ can skip a helpful low-rank warm-up and mildly worsen the final value of the loss function.
  \item \textbf{Rank increment $\Delta r$.} $\Delta r=1$ performs poorly due to too many effective restarts; all the other choices presented for this parameter ($2 \leq \Delta r\leq 4$) demonstrate similar behavior and give reasonable values of the loss function (Figure~\ref{fig:hyps}c).
\end{enumerate}

\paragraph*{Recommendations.}
\begin{enumerate}
  \item Use $s$ as 3--5\% of the total iteration budget. For the $2000$ iterations, this corresponds to $s\approx 60$--$100$.
  \item Do not take $r_{\min}=1$; prefer $r_{\min} \geq 2$ for a stable warm-up without wasting computational time. 
  \item Do not use $\Delta r=1$; we suggest using $\Delta r \simeq r_{\min}$.
\end{enumerate}

\begin{figure*}[!ht]
  \centering
  \begin{minipage}[t]{0.33\linewidth}
    \centering
    \includegraphics[width=\linewidth]{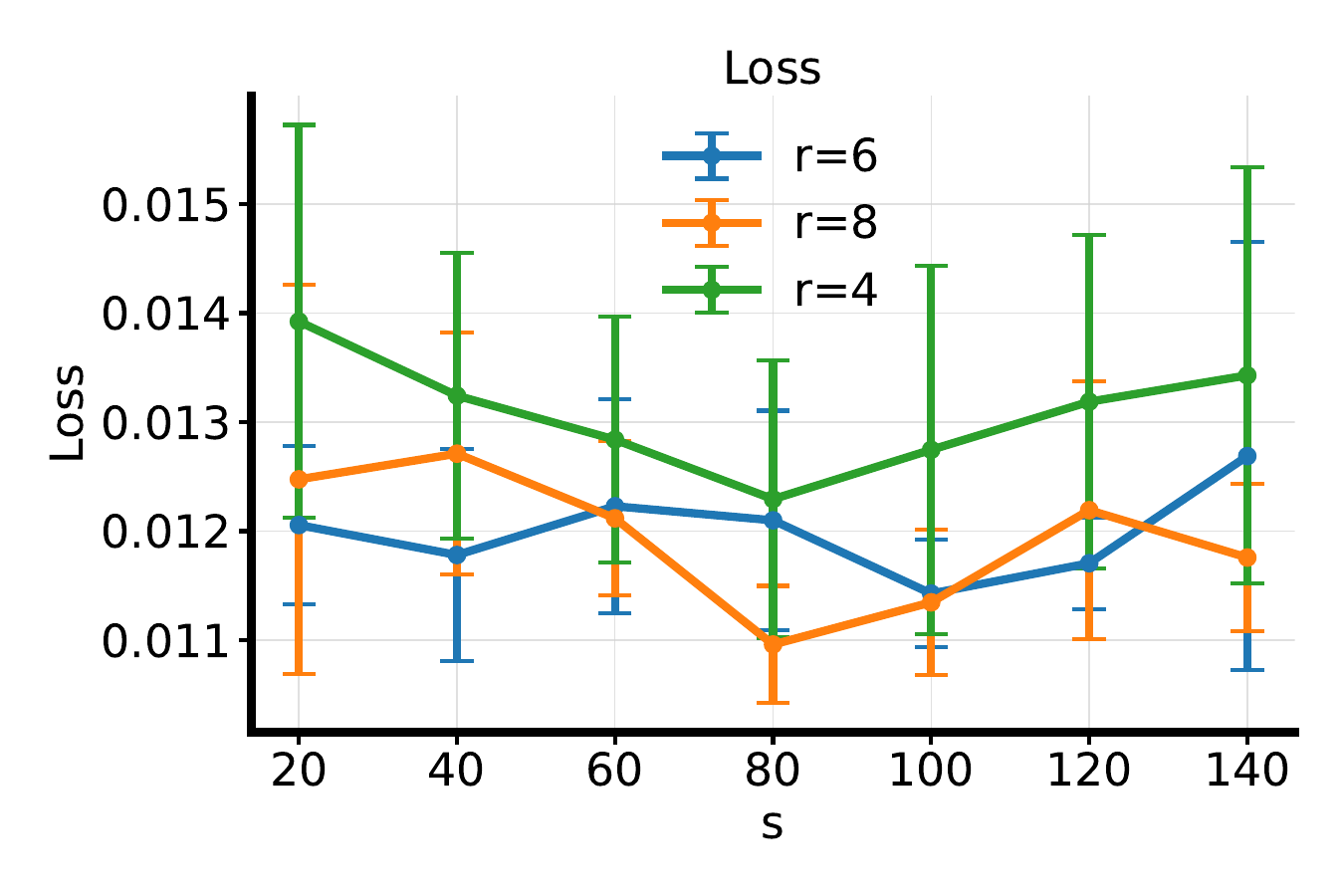}\\[-1pt]
    \small (a) Augmentation interval $s$.
  \end{minipage}\hfill
  \begin{minipage}[t]{0.33\linewidth}
    \centering
    \includegraphics[width=\linewidth]{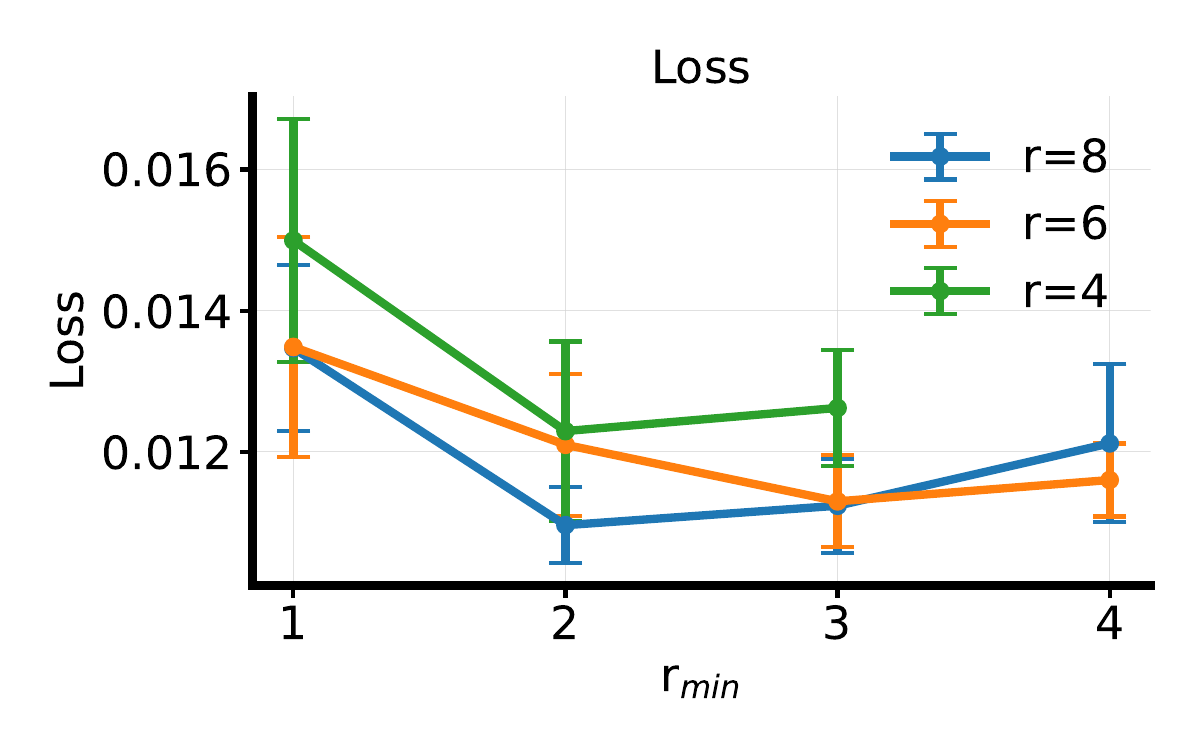}\\[-1pt]
    \small (b) Initial rank $r_{\min}$.
  \end{minipage}\hfill
  \begin{minipage}[t]{0.33\linewidth}
    \centering
    \includegraphics[width=\linewidth]{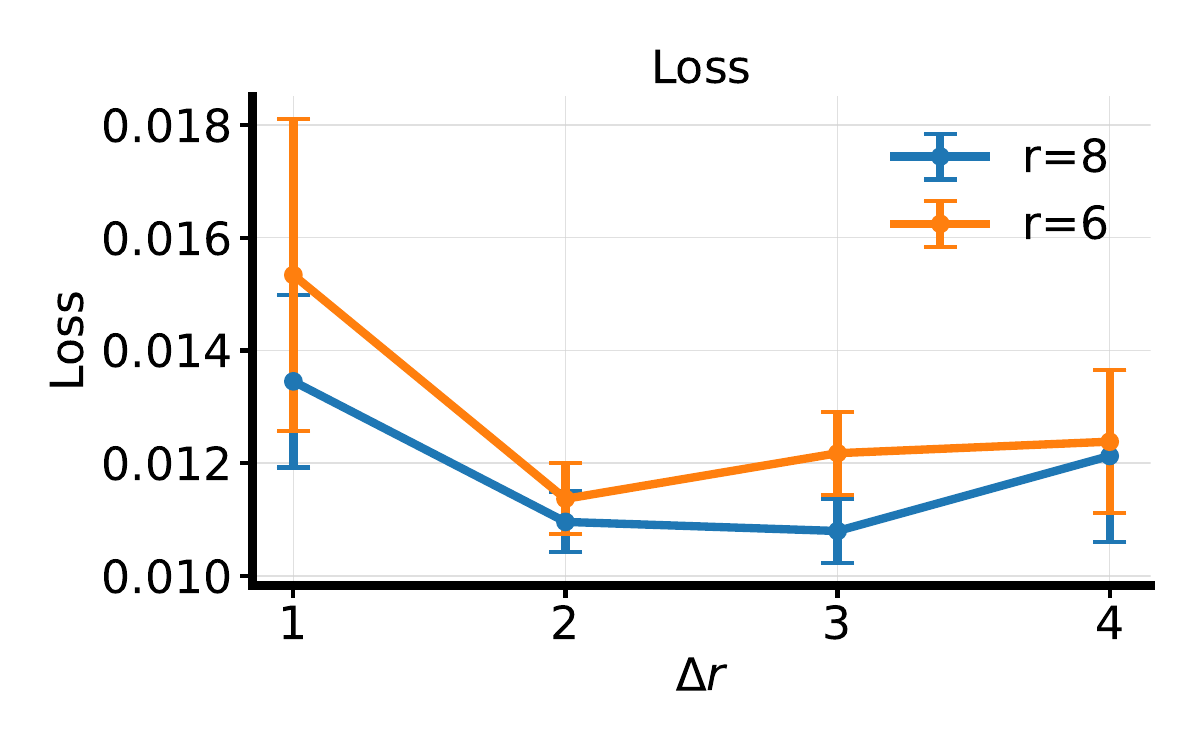}\\[-1pt]
    \small (c) Rank increment $\Delta r$.
  \end{minipage}
  \caption{Dependence of the loss function on hyperparameters for 14th-level MFRA MTPs on the Mo–Nb–Ta–W validation set: (a) the augmentation interval $s$, (b) the initial rank $r_{\min}$, and (c) the rank increment $\Delta r$. Error bars indicate 1-$\sigma$ confidence intervals.}
  \label{fig:hyps}
\end{figure*}

\clearpage
\end{widetext}

\end{document}

%% file: retraction.tikz
\begin{tikzpicture}[scale=4,
  x={(0.8cm,0.2cm)}, y={(1cm,0cm)}, z={(0cm,1cm)},
  declare function={
    f(\x,\y) = 1 - 2*\x*\x - \y*\y;
  }]

% ---- Smooth manifold surface with patches ----
\begin{scope}
  \foreach \x in {-0.3,-0.26,...,0.5} {
    \foreach \y in {-0.8,-0.76,...,0.3} {
      \pgfmathsetmacro{\xA}{\x}
      \pgfmathsetmacro{\yA}{\y}
      \pgfmathsetmacro{\zA}{f(\xA,\yA)}

      \pgfmathsetmacro{\xB}{\x+0.04}
      \pgfmathsetmacro{\yB}{\y}
      \pgfmathsetmacro{\zB}{f(\xB,\yB)}

      \pgfmathsetmacro{\xC}{\x+0.04}
      \pgfmathsetmacro{\yC}{\y+0.04}
      \pgfmathsetmacro{\zC}{f(\xC,\yC)}

      \pgfmathsetmacro{\xD}{\x}
      \pgfmathsetmacro{\yD}{\y+0.04}
      \pgfmathsetmacro{\zD}{f(\xD,\yD)}

      \fill[black!50, opacity=0.4] 
        (\xA,\yA,\zA) -- (\xB,\yB,\zB) -- 
        (\xC,\yC,\zC) -- (\xD,\yD,\zD) -- cycle;
    }
  }
\end{scope}

% --- Point x on the manifold ---
\def\xpt{-0.05}
\def\ypt{0.05}
\pgfmathsetmacro{\zpt}{f(\xpt,\ypt)}
\coordinate (x) at (\xpt,\ypt,\zpt);
\filldraw[black] (x) circle (0.5pt) node[above right=-1pt] {\footnotesize $x$};

% --- Tangent plane ---
\pgfmathsetmacro{\dzdx}{-4*\xpt}
\pgfmathsetmacro{\dzdy}{-2*\ypt}
\coordinate (u) at (1, 0, \dzdx);
\coordinate (v) at (0, 1, \dzdy);

\coordinate (tp1) at ($(x) + 0.4*(u) + 0.4*(v)$);
\coordinate (tp2) at ($(x) - 0.4*(u) + 0.4*(v)$);
\coordinate (tp3) at ($(x) - 0.4*(u) - 0.4*(v)$);
\coordinate (tp4) at ($(x) + 0.4*(u) - 0.4*(v)$);

\fill[orange!90, opacity=0.6] (tp1) -- (tp2) -- (tp3) -- (tp4) -- cycle;
% \draw[black!60!black] (tp1) -- (tp2) -- (tp3) -- (tp4) -- cycle;
\node[orange!50!orange] at ($(x)+(0.5,0.4,0.05)$) {\scriptsize $T_x\mathcal{M}$};

% --- Tangent vector ξ ---
\def\dzdxcoef{-0.22}
\def\dzdycoef{-0.18}
\coordinate (xi) at ($(x) + \dzdxcoef*(1,0,\dzdx) + \dzdycoef*(0,1,\dzdy)$);
\draw[->, thick, black!80] (x) -- (xi) node[midway, above left=-2pt] {\scriptsize $\xi$};

% --- Retraction point R(x,ξ) ---
\pgfmathsetmacro{\xr}{\xpt + \dzdxcoef}
\pgfmathsetmacro{\yr}{\ypt + \dzdycoef}
\pgfmathsetmacro{\zr}{f(\xr,\yr)}
\coordinate (Rx) at (\xr,\yr,\zr);
\filldraw[orange!80!black] (Rx) circle (0.5pt) node[below left=-2pt] {\scriptsize $R(x,\xi)$};

% --- Retraction arc ---
\draw[thick, orange!70!black, densely dashed, ->] 
  plot [smooth, tension=1] coordinates {(xi) ($(xi)!0.5!(Rx)+(0,0,0.05)$) (Rx)};

% --- Label manifold ---
\node at (0,-0.1,0.6) {\scriptsize $\mathcal{M}$};

\end{tikzpicture}

%% file: transport.tikz
\begin{tikzpicture}[scale=4,
  x={(0.8cm,0.2cm)}, y={(1cm,0cm)}, z={(0cm,1cm)},
  declare function={
    f(\x,\y) = 1 - 2*\x*\x - \y*\y;
  }]

% ---- Smooth manifold surface with patches ----
\begin{scope}
  \foreach \x in {-0.3,-0.26,...,0.5} {
    \foreach \y in {-0.8,-0.76,...,0.3} {
      \pgfmathsetmacro{\xA}{\x}
      \pgfmathsetmacro{\yA}{\y}
      \pgfmathsetmacro{\zA}{f(\xA,\yA)}

      \pgfmathsetmacro{\xB}{\x+0.04}
      \pgfmathsetmacro{\yB}{\y}
      \pgfmathsetmacro{\zB}{f(\xB,\yB)}

      \pgfmathsetmacro{\xC}{\x+0.04}
      \pgfmathsetmacro{\yC}{\y+0.04}
      \pgfmathsetmacro{\zC}{f(\xC,\yC)}

      \pgfmathsetmacro{\xD}{\x}
      \pgfmathsetmacro{\yD}{\y+0.04}
      \pgfmathsetmacro{\zD}{f(\xD,\yD)}

      \fill[black!50, opacity=0.4] 
        (\xA,\yA,\zA) -- (\xB,\yB,\zB) -- 
        (\xC,\yC,\zC) -- (\xD,\yD,\zD) -- cycle;
    }
  }
\end{scope}

% --- Point x on the manifold ---
\def\xpt{-0.05}
\def\ypt{0.05}
\pgfmathsetmacro{\zpt}{f(\xpt,\ypt)}
\coordinate (x) at (\xpt,\ypt,\zpt);
\filldraw[black] (x) circle (0.5pt) node[above right=-1pt] {\footnotesize $x$};

% --- Tangent plane ---
\pgfmathsetmacro{\dzdx}{-4*\xpt}
\pgfmathsetmacro{\dzdy}{-2*\ypt}
\coordinate (u) at (1, 0, \dzdx);
\coordinate (v) at (0, 1, \dzdy);

\coordinate (tp1) at ($(x) + 0.4*(u) + 0.4*(v)$);
\coordinate (tp2) at ($(x) - 0.4*(u) + 0.4*(v)$);
\coordinate (tp3) at ($(x) - 0.4*(u) - 0.4*(v)$);
\coordinate (tp4) at ($(x) + 0.4*(u) - 0.4*(v)$);

\fill[orange!90, opacity=0.6] (tp1) -- (tp2) -- (tp3) -- (tp4) -- cycle;
% \draw[green!60!black] (tp1) -- (tp2) -- (tp3) -- (tp4) -- cycle;
\node[orange!50!orange] at ($(x)+(0.5,0.4,0.05)$) {\scriptsize $T_x\mathcal{M}$};

% --- Tangent vector ξ ---
\def\dzdxcoef{-0.22}
\def\dzdycoef{-0.18}
\coordinate (xi) at ($(x) + \dzdxcoef*(1,0,\dzdx) + \dzdycoef*(0,1,\dzdy)$);
\draw[->, thick, black!80] (x) -- (xi) node[midway, above left=-2pt] {\scriptsize $\xi$};

% --- Point y ---
\pgfmathsetmacro{\xr}{\xpt + 1.4*\dzdxcoef}
\pgfmathsetmacro{\yr}{\ypt + 1.4*\dzdycoef}
\pgfmathsetmacro{\zr}{f(\xr,\yr)}
\coordinate (Rx) at (\xr,\yr,\zr);
\filldraw[black!80!blue] (Rx) circle (0.5pt) node[below right=1pt] {\scriptsize $y$};

% --- Tangent plane for retracted point ---
\pgfmathsetmacro{\dzdxr}{-4*\xr}
\pgfmathsetmacro{\dzdyr}{-2*\yr}
\coordinate (u) at (1, 0, \dzdxr);
\coordinate (v) at (0, 1, \dzdyr);

\coordinate (tp1) at ($(Rx) + 0.15*(u) + 0.3*(v)$);
\coordinate (tp2) at ($(Rx) - 0.3*(u) + 0.3*(v)$);
\coordinate (tp3) at ($(Rx) - 0.15*(u) - 0.3*(v)$);
\coordinate (tp4) at ($(Rx) + 0.3*(u) - 0.3*(v)$);

\fill[blue!40, opacity=0.6] (tp1) -- (tp2) -- (tp3) -- (tp4) -- cycle;
% \draw[black!60!black] (tp1) -- (tp2) -- (tp3) -- (tp4) -- cycle;
\node[blue!50!blue] at ($(Rx)+(-1,0.3,0.05)$) {\scriptsize $T_y\mathcal{M}$};

% % --- Retraction arc ---
% \draw[thick, red!70!black, densely dashed, ->] 
%   plot [smooth, tension=1] coordinates {(xi) ($(xi)!0.5!(Rx)+(0,0,0.05)$) (Rx)};

% --- Tangent vector ξ ---
\coordinate (xi2) at ($(Rx) + \dzdxcoef*0.4*(1,0.4,\dzdxr) + \dzdycoef*0.4*(0,1,\dzdyr)$);
\draw[->, thick, black!80] (Rx) -- (xi2) node[midway, above left=-2pt] {\scriptsize $\mathcal{T}_{x\to y}\xi$};

% --- Label manifold ---
\node at (0,-0.1,0.6) {\scriptsize $\mathcal{M}$};

\draw[dashed, thick, black!80]
  plot [smooth, tension=1] coordinates {
    (x)
    ($(x)!0.5!(Rx)+(0,0,0.05)$) % небольшое приподнятие
    (Rx)
  };

\end{tikzpicture}